\documentclass[superscriptaddress,secnumarabic,twocolumn,
 amssymb,amsmath,nobibnotes,aps,prd,showkeys,showpacs,nofootinbib]{revtex4}
\usepackage{amsmath}
\usepackage{amssymb}
\usepackage{graphicx}
\usepackage{color}

\begin{document}

\title{Classes of non-minimally coupled scalar fields in spatially curved FRW spacetimes}
\author{Morteza Kerachian}
\email{morteza.kerachian@gmail.com}
\affiliation{Institute of Theoretical Physics, Faculty of Mathematics and Physics,
Charles University, CZ-180 00 Prague, Czech Republic}
\author{Giovanni Acquaviva}
\email{gioacqua@gmail.com}
\affiliation{Institute of Theoretical Physics, Faculty of Mathematics and Physics,
Charles University, CZ-180 00 Prague, Czech Republic}
\author{Georgios Lukes-Gerakopoulos}
\email{gglukes@gmail.com}
\affiliation{Astronomical Institute of the Academy of Sciences of the Czech Republic,
Bo\v{c}n\'{i} II 1401/1a, CZ-141 00 Prague, Czech Republic}

\begin{abstract}
    In this work we perform a dynamical analysis of a broad class of non-minimally coupled real scalar fields in the Friedmann-Robertson-Walker (FRW) spacetime framework. The first part of our study concerns the dynamics of an unspecified positive potential in a spatially curved FRW spacetime, for which we define a new set of dimensionless variables and a new evolution parameter. In the framework of this general setup we have recognized several general features of the system, like symmetries, invariant subsets and critical points, and provide their cosmological interpretation. The second part of our work focuses on flat FRW cases for which the tracker parameter is constant, i.e. we examine specific classes of potentials. After analyzing these cases dynamically, we discuss their physical interpretation.  
\end{abstract}

\pacs{~~}
\maketitle

\section{Introduction}

The importance of scalar fields in cosmological frameworks is manifold.  In the context of inflation \cite{starobinsky1979aa,guth1981ah,linde1982new,akrami2018planck}, for instance, field theories have been proposed which could appropriately explain the observational evidence of large-scale homogeneity and flatness of the Universe, together with several other features (the graceful exit from inflation itself \cite{albrecht1982albrecht} and the subsequent reheating \cite{kofman1994}).  While the specific mechanism giving rise to such {\it inflaton field} is still debated, several forms of potentials that are able to trigger a transient phase of exponential expansion of the Universe have been proposed, see {\it e.g.} \cite{kolb1990early,galante2015unity}.  Scalar fields play a major role as well in the description of the present-day accelerated expansion of the Universe \cite{aghanim2018planck}: the simplest and most effective model available, the $\Lambda CDM$, considers a constant potential, but the origin of such {\it cosmological constant} is purely phenomenological and cannot be physically motivated in the context of GR and quantum field theory yet; however, other potential forms \cite{zlatev1999quintessence,copeland2006dynamics} are also able to provide the necessary slow-rolling dynamic of the field, which is necessary for achieving a sufficiently negative pressure and consequently an acceleration of the scale factor's expansion.

In a Lagrangian formulation of a gravitating scalar field, the simplest choice is to ignore any direct coupling between the field and the Ricci curvature, {\it i.e.} to consider the so-called {\it minimal coupling}.  However, the inclusion of coupling terms involving products of the Ricci scalar with the field (or its derivatives \cite{amendola1993cosmology,sushkov2009exact,saridakis2010quintessence}) can be motivated in different contexts: they can arise from quantum corrections to the field in curved metrics \cite{allen1983phase,birrelquantum} or as low-energy limits of superstring theories or in induced gravity \cite{maeda1986stability,accetta1985induced}; moreover, a non-minimal coupling can render the Higgs field a good candidate for inflation \cite{bezrukov2008standard}, hence giving a cosmological status to the recently-discovered particle \cite{aad2012g}.  On a more fundamental level, requiring a non-minimal coupling is actually necessary in order to avoid causal pathologies in the propagation of the fields in generic curved backgrounds \cite{faraoni2013conformally}.  Several authors have analysed the repercussions of non-minimal couplings on the cosmological dynamics \cite{barroso1992inflation, uzan1999cosmological, gunzig2000dynamical, bertolami2000nonminimal, de2000tracker, riazuelo2002cosmological,Hrycyna2010, sami2012cosmological, kamenshchik2014integrable, skugoreva2014global}, 

In the present paper we perform a global analysis of models in which a curved Friedmann-Robertson-Walker background is non-minimally coupled to a scalar field with generic potential.  A similar analysis in the context of dynamical systems has been performed in \cite{Hrycyna2010} with the additional presence of matter.  Our goal here is to present an alternative formulation which allows for several improvements in the aforementioned analysis.  Namely we consider a generic spatially curved FRW model and we include in the analysis the collapsing scenarios as well.  In Sec.~\ref{sec:sys} we provide definitions of dimensionless variables that render the invariant subsets compact in a physically relevant range of the coupling parameter $\xi$, without the need of further compactification through an additional change of variables.  In Sec.\ref{sec:CritPoi} we perform an initial analysis keeping the potential of the field completely unspecified (apart from its positivity): this approach covers a class of potentials broader than the ones in \cite{Hrycyna2010}.  Under our general assumptions we derive the existence, stability and cosmological meaning of the critical points of the system.  It is known that the system cannot be closed without specifying the functional form of the potential: in Sec.\ref{sec:specpot} we briefly review the case of exponential potentials and then introduce the analysis of the wide class of potentials characterised by a constant tracker parameter $\Gamma$.  In the latter case, we show that the models with $\Gamma\geq\textrm{const}>1$ and constant always posses de Sitter attractors, irrespective of the value of the other parameters involved.

We start by considering the effective Lagrangian describing a scalar field $\psi$ with generic potential $V(\psi)$ and non-minimally coupled to a FRW background spacetime:
\begin{equation} \label{eq:Lag_FRW_scalar}
 L = 6\, \left(\dot{a}^2 - k \right)\, a\, U(\psi) + 6\, \dot{a}\, a^2\, \dot{\psi}\, U'(\psi) - \frac{1}{2}\, a^3\, \dot{\psi}^2 + a^3\, V(\psi) ,
\end{equation}
where dot and prime denote derivatives w.r.t. the cosmic time and the scalar field respectively.  The function $U(\psi)$ specifies the type of coupling considered: minimal coupling corresponds to a constant $U=1/2$, while in the following we will consider the quadratic form
\begin{equation}\label{eq:coupl}
 U = \frac{1}{2}\, \left( 1 - \xi\, \psi^2 \right)\, ,
\end{equation}
with $\xi\geq0$. The case $\xi=1/6$ corresponds to the conformal coupling.
With the choice Eq.~\eqref{eq:coupl}, we can explicitly calculate the momenta conjugate to the generalized coordinates $\{ a, \psi \}$, namely
\begin{align}
 p_a &\equiv \frac{\partial L}{\partial \dot{a}} = 6\, \dot{a}\, a \left(1-\xi\, \psi^2 \right) - 6\, \xi\, a^2\, \psi\, \dot{\psi}\\
 p_{\psi} &\equiv \frac{\partial L}{\partial \dot{\psi}} = -6\, \xi\, a^2\, \dot{a}\, \psi - a^3\, \dot{\psi}\, ,
\end{align}
and hence the Hamiltonian function
\begin{align}
 \mathcal{H} &\equiv p_a\, \dot{a}+ p_{\psi}\, \dot{\psi} - L
\end{align}
The Hamiltonian constraint is expressed by the condition $\mathcal{H}=0$ and it corresponds to Friedmann equation
\begin{equation} \label{eq:fried}
 3\, \left( H^2 + \frac{k}{a^2} \right)\, \left( 1- \xi\, \psi^2 \right) = 6\, \xi\, H\, \psi\, \dot{\psi} + \frac{1}{2}\, \dot{\psi}^2 + V(\psi)\, ,
\end{equation}
where $H=\dot{a}/a$ is the FRW Hubble expansion.  The Hamilton-Jacobi equations,
\begin{equation}
 \dot{p}_{a} = \frac{\partial L}{\partial a}\quad ,\quad \dot{p}_{\psi} = \frac{\partial L}{\partial \psi}\, ,
\end{equation}
correspond, respectively, to Raychaudhuri and Klein-Gordon equations:
\begin{align}
 &\left( 2\,\dot{H} +3\, H^2 + \frac{k}{a^2} \right) \left(1-\xi\, \psi^2\right) - 4\, \xi\, H\, \psi\, \dot{\psi} - 2\, \xi\, \psi\, \ddot{\psi} = \nonumber \\
 &- \left(1-4\, \xi\right)\frac{1}{2}\, \dot{\psi}^2 + V(\psi)\label{eq:ray1}\\
 &\ddot{\psi} + 3\, H\, \dot{\psi} + \partial_{\psi} V + 6\, \xi\, \psi\,  \left(\dot{H} + 2\, H^2 + \frac{k}{a^2} \right) = 0\, .\label{eq:klein1}
\end{align}

\section{The system in a new set of variables}
\label{sec:sys}

In the minimally coupled case one can clearly distinguish two behaviours of the dynamics depending on the sign of the spatial curvature: specifically, if $k>0$ the expansion scalar can change sign during the evolution, leading to bounces or recollapses, while if $k\leq0$ the solutions are either always expanding or always contracting.  For this reason, in order to construct well-defined dimensionless variables in the case of positive curvature, one usually employs the normalization $\sqrt{H^2 + k/a^2}$ which is positive definite and does not vanish at the turning points of the scale factor.  Introducing a nonminimal coupling renders the former distinction meaningless, due to the modifications of the Raychaudhuri equation which allow for sign changes of $H$ during the evolution irrespective of the sign of $k$.  Since now the evolution of the scale factor can present turning points in either curvature cases, we define a set of dimensionless variables which is well-defined for both:
\begin{align}
 \Omega = \frac{\psi}{\sqrt{1+\xi\, \psi^2}}\quad &,\quad \Omega_H = \frac{H}{D}\label{var1}\\
 \Omega_{\psi} = \frac{\dot{\psi}}{\sqrt{6}\, D}\quad &,\quad \Omega_V = \frac{\sqrt{V}}{\sqrt{3}\, D}\\
 \Omega_{\partial V} = \frac{\partial_\psi V}{V}\quad &,\quad \Gamma = \frac{V \cdot \partial^2_\psi V}{(\partial_\psi V)^2}
 \label{eq:var2}
\end{align}
where
\begin{equation}
D^2=H^2+\frac{|k|}{a^2}\, .
\end{equation}
A useful relation is the time evolution of $D$ in terms of the dimensionless variables:
\begin{equation}\label{ddot}
 \frac{\dot{D}}{D^2} = \Omega_H\, \left( \frac{\dot{H}}{D^2} + \Omega_H^2 - 1 \right)\, .
 \end{equation}
The Friedmann, Raychaudhuri and Klein-Gordon equations in terms of the normalized variables will take a different form depending on the sign of the spatial curvature (see next subsections).  It is however possible to derive a common autonomous system of equations for the variables, with evolution parameter defined by $d\tau = D\, dt$, by taking derivatives of the definitions with respect to such parameter and using Eq.~\eqref{ddot}:
\begin{align}
 \Omega' &= \sqrt{6}\ \Omega_\psi\ \left( 1-\xi\, \Omega^2 \right)^{3/2} \label{eq:omega}\\
 \Omega_H' &= \left( 1 - \Omega_H^2 \right)\, \left( \frac{\dot{H}}{D^2} + \Omega_H^2 \right) \label{eq:omegah}\\
 \Omega_\psi' &= \frac{\ddot{\psi}}{\sqrt{6}\, D^2} - \Omega_\psi\, \Omega_H\, \left( \frac{\dot{H}}{D^2} + \Omega_H^2 - 1 \right) \label{eq:omegapsi}\\
 \Omega_V' &= \Omega_V\, \left[ \sqrt{\frac{3}{2}}\ \Omega_{\partial V}\ \Omega_{\psi} - \Omega_H\, \left( \frac{\dot{H}}{D^2} + \Omega_H^2 - 1 \right) \right] \label{eq:omegav}\\
 \Omega_{\partial V}' &= \sqrt{6}\ \Omega_{\partial V}^2\ \Omega_\psi\, \left( \Gamma - 1 \right)\, , \label{eq:omegadv}
\end{align}
where $\Gamma=V\, \cdot\, \partial^2_{\psi} V / \left( \partial_{\psi} V \right)^2$ is the so-called tracker parameter.  The quantities $\dot{H}$ and $\ddot{\psi}$ are obtained by decoupling Eq.~\eqref{eq:ray1} and Eq.~\eqref{eq:klein1} and they determine different dynamics for the two curvature cases. For the generic non-minimally coupled cases the decoupling of the Eqs.~\eqref{eq:ray1},~\eqref{eq:klein1} can be achieved by diagonalizing the following linear system:
\begin{align}\label{eq:decoup}
    \begin{bmatrix}
    2(1-\xi \psi^2) & -2\xi \psi \\
    6\xi \psi & 1\\
    \end{bmatrix}
 \begin{bmatrix}
  \dot{H} \\ \ddot{\psi}
   \end{bmatrix}
   =
 \begin{bmatrix}
  f_1(\Omega_i) \\ f_2(\Omega_i)
   \end{bmatrix},
\end{align}
where $f_1(\Omega_i)$ and $f_2(\Omega_i)$ include the terms which are not linear in $\dot{H}$ and $\ddot{\psi}$ in Raychaudhuri and Klein-Gordon equations respectively, with $\Omega_i$ representing the set of dimensionless variables. In order to diagonalize the matrix in Eq.~\eqref{eq:decoup} its determinant should be non-zero, i.e. $\psi^2\, \xi\, (1-6\xi)~\neq~1$.  The case $\xi=0$ is trivial, while the conformal coupling case $\xi=1/6$, as we will see, leads to a generic unboundedness of the invariant subsets of the system.  We will be mostly interested in the range $\xi\in(0,1/6)$ for two reasons: first of all, the invariant subsets of the system in this range of the parameter are compact; moreover, in \cite{hrycyna2017xi} the value of the coupling constant has been constrained using observational data from the {\it Union2.1+H(z)+Alcock-Paczy\'{n}ski} data set and found to be in good accord with values around the conformally coupled case.  In this sense we will scan the behaviour of the system inside the intersection between the physically motivated and the mathematically convenient range.  The vanishing of the determinant for specific values of the field implies the appearance of singularities in the system.  Such anomalies are independent of the definition adopted for the dimensionless variables: different definitions would simply move the singularities in different parts of the parameter space.  We point out that our choice of dimensionless variables is particularly suitable for the analysis of the late-time behaviour of the system and for situations in which the scalar field diverges $\psi\rightarrow\pm \infty$, because the new variable $\Omega$ remains finite.

\subsection{Positive curvature}

When $k>0$, the Friedmann equation can be expressed in terms of the variables Eqs.~\eqref{var1}-\eqref{eq:var2} in the following form:
\begin{align}
 1 =\ &2\, \xi\, \Omega^2\, \left( 1 - \Omega_H^2 \right) + 3\, \xi\, \left( \sqrt{\frac{2}{3}}\, \Omega_H\, \Omega + \Omega_{\psi}\, \sqrt{1-\xi\, \Omega^2} \right)^2 \nonumber\\ 
 &+ (1-3\, \xi)\, \Omega_{\psi}^2\, \left(1-\xi\, \Omega^2\right) + \Omega_V^2\, \left(1-\xi\, \Omega^2\right)\label{eq:fried_pos}
 \end{align}
Since from the definitions we have that $\Omega\in\left( -1/\sqrt{\xi} , 1/\sqrt{\xi} \right)$ and $\Omega_H\in\left( -1, 1 \right)$, the constraint Eq.~\eqref{eq:fried_pos} defines a compact parameter space if $\xi\in(0,1/6)$ (see discussion in Sec.~\ref{sec:GenFeat} paragraph b).  From Klein-Gordon and Raychaudhuri equations we get
\begin{align}
\frac{\ddot{\psi}}{\sqrt{6}\ D^2} &= - 3\ \Omega_H\ \Omega_\psi - \sqrt{\frac{3}{2}}\ \Omega_{\partial V}\ \Omega_V^2 \nonumber \\
 &- \frac{\sqrt{6}\ \xi\ \Omega}{\sqrt{1-\xi\ \Omega^2}} \left(  \frac{\dot{H}}{D^2} + \Omega_H^2 + 1 \right)\, \label{eq:DeKlPos} ,\\
 \frac{\dot{H}}{D^2} + \Omega_H^2 + 1 &= - \frac{1}{1-2\ \xi\ (1-3\, \xi)\ \Omega^2}\ \Bigg\{-\frac{1}{2}\left( 1-2\, \xi\, \Omega^2 \right) \nonumber\\ 
 &+ \xi\, \Omega\, \sqrt{1-\xi\, \Omega^2} \left( \sqrt{6}\, \Omega_H\, \Omega_\psi+3\, \Omega_{\partial V}\, \Omega_V^2 \right)\nonumber \\ 
 &+ \frac{3}{2}\left(1-\xi\, \Omega^2\right) \Big[ (1-4\, \xi)\, \Omega_\psi^2-\Omega_V^2 \Big]  \Bigg\}
 \label{eq:DeRaychPos}
\end{align}

\subsection{Non-positive curvature}

Applying the same definitions given by Eqs.~\eqref{var1}-\eqref{eq:var2} to the case of non-positive spatial curvature $k\leq0$, one can express the Friedmann constraint in the following form:
\begin{align}
 1 =\ &2\, \left(1-\xi\, \Omega^2\right)\, \left(1-\Omega_H^2\right) \nonumber \\
  &+  3\, \xi\, \left( \sqrt{\frac{2}{3}}\, \Omega_H\, \Omega + \Omega_{\psi}\, \sqrt{1-\xi\, \Omega^2} \right)^2\nonumber\\ 
 &+ (1-3\, \xi)\, \Omega_{\psi}^2\, \left(1-\xi\, \Omega^2\right) + \Omega_V^2\, \left(1-\xi\, \Omega^2\right)\, .\label{eq:fried_neg}
\end{align}
In this case the parameter space spanned by such variables is not compact, because $\Omega_\psi$ diverges as $\Omega\rightarrow\pm1/\sqrt{\xi}$. Eqs.~\eqref{eq:ray1}-\eqref{eq:klein1} give
\begin{align}
\frac{\ddot{\psi}}{\sqrt{6}\ D^2} = &- 3\ \Omega_H\ \Omega_\psi - \sqrt{\frac{3}{2}}\ \Omega_{\partial V}\ \Omega_V^2 \nonumber \\
 &+ \frac{\sqrt{6}\ \xi\ \Omega}{\sqrt{1-\xi\ \Omega^2}} \left(  1- \frac{\dot{H}}{D^2} - 3\, \Omega_H^2 \right)\, \label{eq:DeKlNPos}  ,\\
 \frac{\dot{H}}{D^2} +\Omega_H^2 &= \frac{1}{2} - \Omega_H^2 \nonumber \\
 &+ \frac{1}{1-2\ \xi\ (1-3\, \xi)\ \Omega^2}\ \Bigg\{ 3\, \xi^2\, \Omega^2\, \left( 1-2\, \Omega_H^2 \right)\nonumber \\ 
 &- \xi\, \Omega\, \sqrt{1-\xi\, \Omega^2} \left( \sqrt{6}\, \Omega_H\, \Omega_\psi+3\, \Omega_{\partial V}\, \Omega_V^2 \right)\nonumber \\ 
 &- \frac{3}{2}\left(1-\xi\, \Omega^2\right) \Big[ (1-4\, \xi)\, \Omega_\psi^2-\Omega_V^2 \Big] \Bigg\}\, 
  \label{eq:DeRaychNPos}.
\end{align}

\subsection{General features of the system} \label{sec:GenFeat}
\paragraph{Symmetries.}
The dynamical system (\ref{eq:omega})-(\ref{eq:omegadv}) remains invariant under the simultaneous transformation $\{\Omega, \Omega_{H}, \Omega_{\psi}, \Omega_{V}, \Omega_{\partial V} \} \to \{-\Omega,\Omega_{H},-\Omega_{\psi},\Omega_{V},-\Omega_{\partial V} \}  $.  Physically such symmetry is equivalent to the invariance under the transformation $\psi \rightarrow -\psi$.  Having assumed the positivity of the potential, we have that $V(-\psi)$ is still positive and hence $\Omega_V$ is not affected by this transformation.

\paragraph{Singularities.}
As we have discussed before, the decoupling of Raychaudhuri and Klein-Gordon equations cannot be carried out if the determinant of Eq.~\eqref{eq:decoup} vanishes: the points where this is the case appear as singularities in the autonomous system.  In terms of dimensionless variables these singularities correspond to the vanishing of the denominators in Eqs.~\eqref{eq:DeRaychPos} and~\eqref{eq:DeRaychNPos}, namely 
\begin{equation} \label{eq:sing}
    \Omega=\pm \frac{1}{\sqrt{2\xi(1-3\xi)}}.
\end{equation}
By plugging Eq.~\eqref{eq:sing} into the Friedmann constraints and solving for $\Omega_{\psi}$ we get
\begin{equation}
    \Omega_{\psi}=\frac{\sqrt{6\xi}\Omega_H+\sqrt{(\Omega_H^2\mp\Omega_V^2-1)6\xi\pm \Omega_V^2}}{\sqrt{1-6\xi}},
\end{equation}
where the upper/lower sign corresponds to negative/positive curvature. In either cases the coordinates $\left( \Omega, \Omega_{\psi} \right)$ of the singularity remain finite in the range $\xi \in (0,1/6)$.  For $\xi>1/6$, $\Omega_\psi$ is complex.
In the case of a flat spacetime $\Omega_H=\pm 1$ we call these singularities $S_{\pm}$ respectively. Comparing with \cite{Hrycyna2010}, we note that for $\Omega_H^2=1$ and $\Omega_V=0$ this corresponds to their critical point 1., which was identified as a finite scale factor singularity. Such critical point was identified thanks to a time reparametrization (see  eq.~(2.14) of \cite{Hrycyna2010}), which, however, we are not considering here as it is ill-defined in the point~\eqref{eq:sing}.

\begin{figure*}
    \centering
        \includegraphics[width=0.3\textwidth]{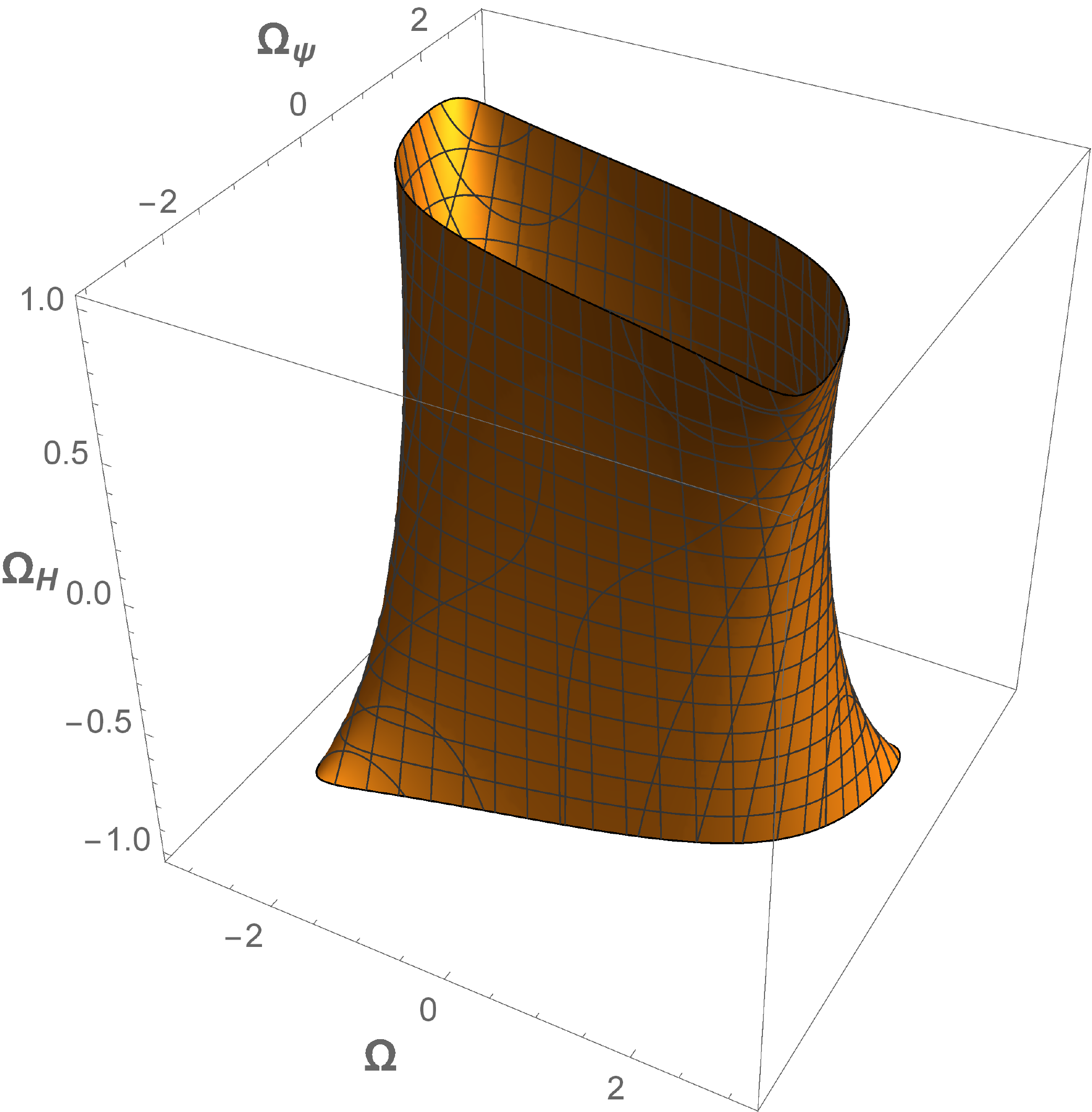}
        \includegraphics[width=0.3\textwidth]{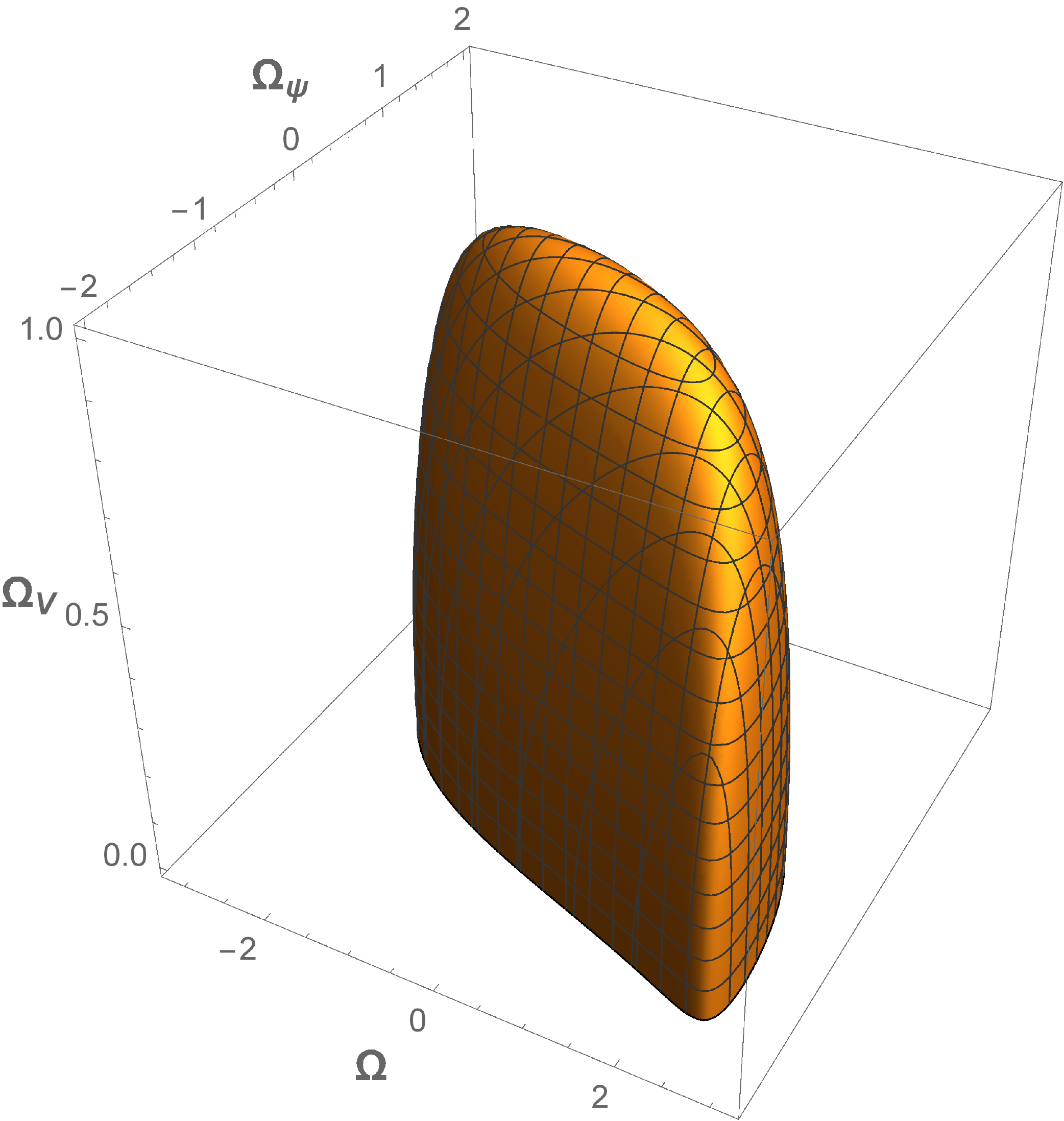}
        \includegraphics[width=0.3\textwidth]{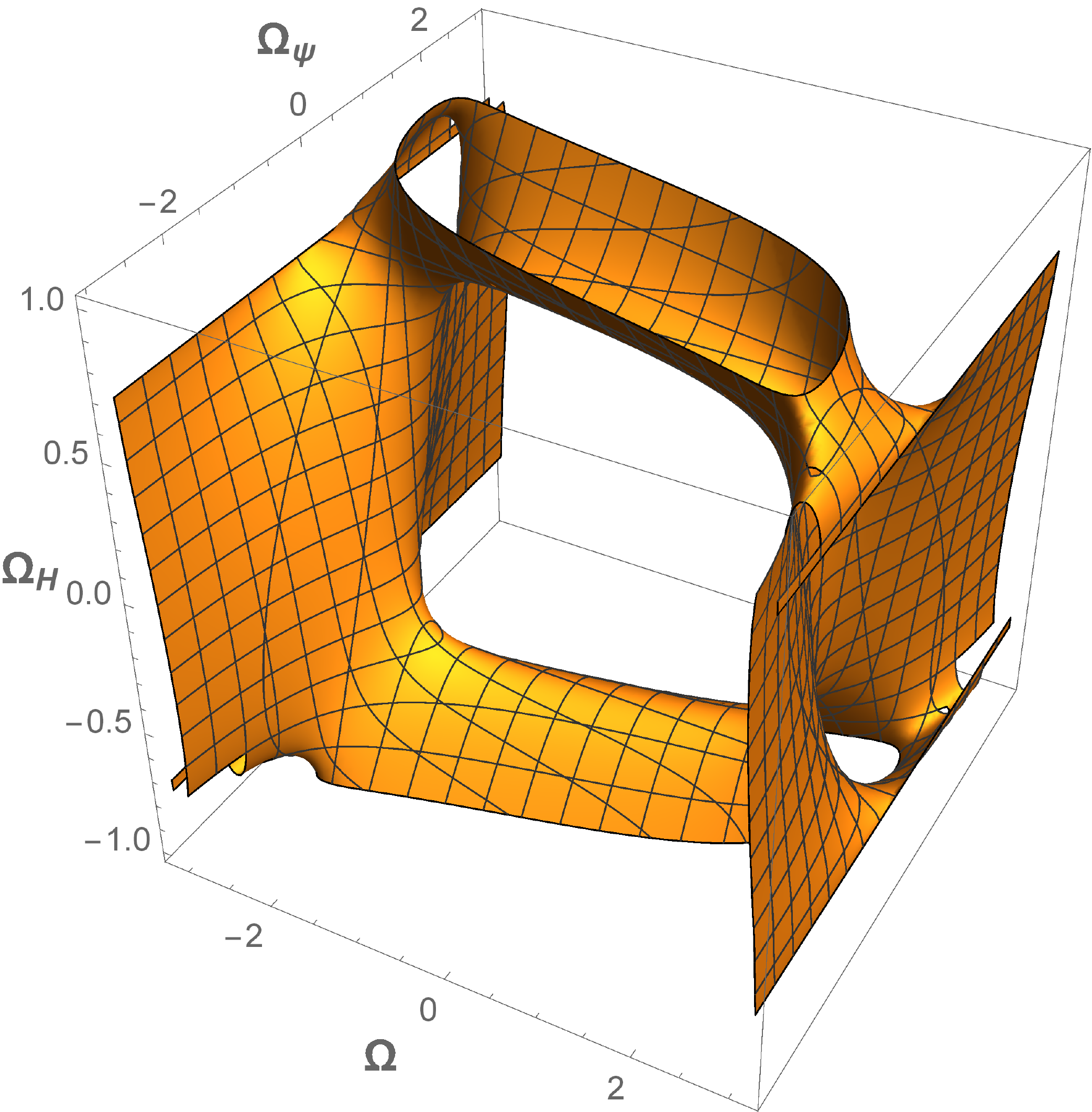}
    \caption{Invariant subsets constrained by the Friedmann equations for $\xi=1/10$. {\it Left Panel:} Positive curvature Eq.~\eqref{eq:fried_pos} for $\Omega_V=0$.  {\it Middle Panel:} Spatially flat Eqs.~\eqref{eq:fried_pos} and \eqref{eq:fried_neg} for $\Omega_H=1$.  {\it Right Panel:} Non-positive curvature Eq.~\eqref{eq:fried_neg} for $\Omega_V=0$.}\label{fig:fried}
\end{figure*}

\paragraph{Invariant subsets.}
Invariant submanifolds are very useful tools in studying a dynamical system, as they allow to characterize and understand some global features of the phase space. One can identify some invariant subsets of the system Eqs.~\eqref{eq:omega}-\eqref{eq:omegadv}, namely $\Omega_H=\pm1$ (flat spacetime) and $\Omega_V=0$ (free scalar field).  For the latter case we plot in the left and right panels of Fig.~\ref{fig:fried} the Friedmann constraints \eqref{eq:fried_pos} and \eqref{eq:fried_neg} respectively; in the middle panel of Fig.~\ref{fig:fried} we plot the Friedmann constraint in the spatially flat expanding case $\Omega_H=1$ (the collapsing case can be obtained by transforming $\Omega_{\psi}\rightarrow-\Omega_{\psi}$).  Notice that our definitions of variables allow to have compact invariant subsets for the positive and zero curvature cases, but not for the negative curvature case.

Although from the system of equations $\Omega=\pm1/\sqrt{\xi}$ looks like an invariant subset, it is actually outside of the Friedmann constraint in the case of positive and zero spatial curvature; for negative curvature, instead, the Friedmann constraint in that locus reduces to $\Omega_H^2=1/2$.

The condition $\Omega_{\partial V}=\textrm{const.}$, which is equivalent to $\Gamma=1$ (including the $\Omega_{\partial V}=0$ case), looks also like an invariant subset due to Eq.~\eqref{eq:omegadv}, but this is a more subtle case, since choosing a constant value of $\Omega_{\partial V}$ actually constraints the form of the potential to the exponential form $V=V_0\, e^{\Omega_{\partial V}\, \psi}$ (see, e.g., \cite{Bahamonde2018} and references therein). We will discuss these kind of potentials in Sec~\ref{sec:specpot}. However, being the potential $V$ a function of the field $\psi$ only, in order to allow for the most general forms of the potential, $\Omega_{\partial V}$ has to be left as a general function of $\Omega$.

\section{Critical points and their interpretation}\label{sec:CritPoi}

\begin{table*}
  \caption{The critical elements of the system and their stability in the range $0\le\xi\le 1/6$.  
  }
\begin{tabular}{c | c c c c c | c c c c}\label{tab:CritPoints}
 & $\Omega_{\psi}$ &  $\Omega_H$ & $\Omega$ & $ \Omega_V$ & $\Omega_{\partial V}$ & Curvature & $q$ & $w_{e}$ & stability\\
\hline
 & & & & &  & & & & \\
  
  $A_+$ & 0 & $1$ & 0 & $ 1$ & 0 & flat & -1 & -1 & sink\\
  $A_-$ & 0 & $-1$ & 0 & $ 1$ & 0 & flat & -1 & -1 & source\\
  $B_+$ & 0 & $1$ & $0<\Omega^2<\frac{1}{2\xi}$ & $ \sqrt{\frac{1-2\xi \Omega^2}{1-\xi \Omega^2}}$ & $-\frac{4\xi\Omega\sqrt{1-\xi\Omega^2}}{1-2\xi\Omega^2}$  & flat & -1 & -1 & sink\\
  $B_-$ & 0 & $-1$ & $0<\Omega^2<\frac{1}{2\xi}$ & $\sqrt{\frac{1-2\xi \Omega^2}{1-\xi \Omega^2}}$ & $-\frac{4\xi\Omega\sqrt{1-\xi\Omega^2}}{1-2\xi\Omega^2}$  & flat & -1 & -1 & source\\
  $C_{\pm}$ & 0 & $\pm 1$ & $\pm\frac{1}{\sqrt{2\xi}}$ & 0 & $\forall$ & flat & 1 & $\frac{1}{3}$ & saddle \\
  $D_{\pm}$ & 0 & $\pm \frac{1}{\sqrt{2}}$ & $\forall$ & 0 & $\forall$  & negative & 0 & - 
     & saddle\\
\end{tabular}
\end{table*}

To study the behaviour of the dynamical system Eqs.~(\ref{eq:omega})-(\ref{eq:omegadv}), we need to derive the equilibrium points of the system. The equilibrium points (or critical points) of the system $\boldsymbol{\Omega}^{\prime}=\boldsymbol{f}(\boldsymbol{\Omega})$ correspond to those points $\boldsymbol{\Omega_{c}}$ that satisfy $\boldsymbol{\Omega}^{\prime}(\boldsymbol{\Omega_{c}})=0$, which means that the system is at rest.  The stability of the critical points can be investigated by inspecting the eigenvalues of the Jacobian matrix of the linearized system evaluated at each critical point: if the real part of all eigenvalues is positive (resp. negative), then the point is an unstable source (resp. stable sink); mixed signs of the eigenvalues signal the presence of a saddle point; the presence of vanishing eigenvalues means that the critical point is non-hyperbolic and one would need to implement further method in order to ascertain the stability unambiguously -- or resort to numerical and visual approaches.

One can interpret the critical points in terms of cosmological models thanks to several physical quantities, such as the {\it deceleration parameter}
\begin{equation}
 q = -1-\Omega_H^{-2}\, \frac{\dot{H}}{D^2}\, ,
\end{equation}
and the {\it effective equation of state parameter}, which stems from considering the scalar field as a barotropic fluid sourcing the unmodified Einstein's equations with equation of state
\begin{equation} \label{eq:EoS}
 w_{e} = \frac{p_{e}}{ \epsilon_{e}},
\end{equation}
where from Eqs.~\eqref{eq:fried},~\eqref{eq:ray1} we define the {\it effective energy density} and {\it pressure}, respectively:
\begin{align}
    \epsilon_{e}&:=3 \left(H^2+\frac{k}{a^2} \right), \label{eq:EfEn} \\
     p_{e}&:=-2 \dot{H}-3 H^2-\frac{k}{a^2}  \label{eq:EfPr}.
\end{align}

\subsection{Two de Sitter critical points}\label{sec:CPA}

The coordinates of these two critical points are $\{\Omega,\Omega_{H},\Omega_\psi,\Omega_{V},\Omega_{\partial V}\}=\{0,\pm 1,0,1,0 \}$.  One of these points (called $A_+$) has $\Omega_{H}=1$ and it describes an exponentially expanding model, i.e $a \sim e^{H_{0}t}$, with the typical cosmological constant behaviour given by $q=-1$ and $w_e=-1$.  The corresponding eigenvalues are
\begin{align}
    \{\lambda^{A_{+}}_{i} \}&=\{-3 ,-2 ,0 ,-\frac{\sqrt{3}}{2} \Big( \sqrt{3}+\sqrt{3-16\xi} \Big), \nonumber \\
    &\frac{\sqrt{3}}{2}\Big(-\sqrt{3}+\sqrt{3-16\xi}\Big) \},\label{eigap}
\end{align}
where the $i=1,...,5$.  The real parts of all the non-vanishing eigenvalues is always negative.  

The critical point with $\Omega_{H}=-1$ (called $A_{-}$) describes an exponentially collapsing model, i.e. $a \sim e^{-H_{0}t}$ with $q=-1$ and $w_e=-1$.  The eigenvalues in this case are 
\begin{align}
     \{\lambda^{A_{-}}_{i} \}&= \{3 ,2 ,0,\frac{\sqrt{3}}{2}\Big(\sqrt{3}-\sqrt{3-16\xi}\Big), \nonumber \\
     &\frac{\sqrt{3}}{2}\Big(\sqrt{3}+\sqrt{3-16\xi}\Big)\}, \label{eigam}
\end{align}
The real part of all the non-vanishing eigenvalues is always positive. 

For both points, the eigenvalues are complex in the range $\xi>3/16$: this signals a transition of the character of the critical points from node to focus and it is in accord with the findings of \cite{hrycyna2015cosmological}.
 Since $\Omega_{\partial V}=0$, then in a neighborhood of the critical points  $V=V_0>0$: this eliminates the relevance of the $\Omega_{\partial V}'$ equation in such neighborhood. Using the remaining $4\times4$ system of equations with $\Omega_{\partial V}=0$, one recovers exactly the above sets of eigenvalues~\eqref{eigap},~\eqref{eigam} where the $\lambda_3^{A_\pm}=0$ are missing. This indicates that indeed the $A_+$ and $A_-$ are a sink and a source respectively.   

\subsection{Two de Sitter critical lines}\label{sec:CPB}
These critical points lie along the segments $0 < \Omega^2 < \frac{1}{2\xi}$ for the cases $\Omega_{H}=\pm 1$ with $\Omega_{\psi}=0$, $\Omega_{V}^2=\frac{1-2\xi \Omega^2}{1-\xi \Omega^2}$ and\footnote{Note that the since $\Omega_{V}>0$ by definition, the only acceptable solution is $\Omega_{V}=\sqrt{\frac{1-2\xi \Omega^2}{1-\xi \Omega^2}}$ (Table~\ref{tab:CritPoints}). Similarly, in Sec.~\ref{sec:CPA} from $\Omega_V=\pm 1$ we accept only $\Omega_V=1$.}
\begin{align} \label{eq:genOmegdv}
\Omega_{\partial V}=-\frac{4\xi\Omega\sqrt{1-\xi\Omega^2}}{1-2\xi\Omega^2} \,.
\end{align}
In this case one can derive a form of the potential in a neighbourhood of the critical lines by integrating the definition of $\Omega_{\partial V}$ as a function of $\Omega$ given above: transforming back to the variable $\psi$ one obtains
\begin{equation} \label{eq:pot}
    V=V_0(1-\xi \psi^2)^2,
\end{equation}
as well as $\displaystyle H=\pm \sqrt{V_0(1-\xi \psi^2)/3}$. Potential~\eqref{eq:pot} has a Higgs-like form which can provide a symmetry breaking Goldstone mechanism.  One can see that $\Omega=0$ corresponds to the local maximum of the potential, while $\Omega=\pm1/\sqrt{2\xi}$ correspond to the global minima.

Exactly on the critical lines, both the potential $V$ and the Hubble parameter $H$ are constant, thus describing exponentially expanding and collapsing models with $a \sim e^{\pm H t}$ respectively. For calculating the eigenvalues below, we need to specify  $\Gamma$. To do this, we use the local expression of the potential~\eqref{eq:pot}. These points for $\Omega_H=1$ (called $B_{+}$) describe sources, since they have eigenvalues 
\begin{equation}
       \{\lambda^{B_{+}}_{i}\}=\{0,-2, 0, -3, -3\},
\end{equation}
which holds in the allowed ranges of $\xi$ and $\Omega$.

The critical points for $\Omega_H=-1$ (called $B_{-}$) have eigenvalues 
\begin{equation}
    \{\lambda^{B_{-}}_{i}\}= \{0,2, 3, 3, 0\},
\end{equation}
thus, we can interpret $B_{-}$ as source points.

Critical points $A_+$ and $B_+$ agree with the critical points 5 of \cite{Hrycyna2010}, in our analysis there are additionally the $A_-$ and $B_-$ sources describing collapsing models. As it is stressed in \cite{Hrycyna2010} the evolution of the system is independent of the form of the potential, but we find that in the neighborhood of $B_\pm$ the potential has to acquire the form~\eqref{eq:pot}.

One would expect that in the limit $\Omega \rightarrow 0$ one should recover the eigenvalues of the previous critical point, {\it i.e.} $\{\lambda^{B_{\pm}}_{i}\}\rightarrow \{\lambda^{A_{\pm}}_{i}\}$, which however is not the case, since potential~\eqref{eq:pot} is just an approximation holding in the neighbourhood of the critical line. However, the feature that matters for the local stability is the sign of the  $\{\lambda^{B_{\pm}}_{i}\}$. Just like in Sec.~\ref{sec:CPA}, specifying the local form of the potential makes one equation of motion redundant and thus reduces the dimensionality of the system.

\subsection{Two radiation-like critical lines}\label{sec:CPC}

There exist other sets of critical points arranged as critical lines with coordinates
\begin{align}
    \{\Omega,\Omega_{H},\Omega_\psi,\Omega_{V},\Omega_{\partial V}\}=
    \{\pm \frac{1}{\sqrt{2\xi}},\pm 1,0,0,\forall \}.
\end{align}
The cosmological parameters at these points are $q=1$ and $w_e=\frac{1}{3}$, being in agreement with the model describing a radiation dominated universe in which the scale factor evolves like $a \sim \sqrt{t}$ . The corresponding eigenvalues are
\begin{equation}
    \{\lambda^{C_{+}}_{i}\}=\{2,2,-1,1,0\},
\end{equation}
for $\Omega_{H}=1$ (called $C_{+}$), and
\begin{equation}
        \{\lambda^{C_{-}}_{i}\}=\{-2,-2,-1,1,0\},
\end{equation}
for $\Omega_{H}=-1$ (called $C_{-}$).

To investigate the exact form of scale factor, from Raychaudhuri equation we get
\begin{equation}
    H=\frac{1}{2 (t-t_0)+\frac{1}{H_0}},
\end{equation}
where $H_0$ is the Hubble parameter value at time $t_0$ with $a_0=1$. For expanding models ($\dot{a}>0$) $a=\sqrt{2 H_0(t-t_0)+1}$ with $t>t_0-\frac{1}{2 H_0}$, while for collapsing ($\dot{a}<0$) $a=\sqrt{-(2 H_0(t-t_0)+1)}$ with $t<t_0-\frac{1}{2 H_0}$.  Since the eigenvalues of both critical lines have real parts with mixed signs, they correspond to saddle points. The set of points $B_+$ agrees with the critical point 3.b of \cite{Hrycyna2010}.

\subsection{Two Milne-like critical planes}\label{sec:CPD} 

These critical points lie on planes defined by $\{\Omega,\Omega_{H},\Omega_{\psi},\Omega_{V},\Omega_{\partial V}\}=\{\forall,\pm \frac{1}{\sqrt{2}},0,0,\forall \}$ .\footnote{Note that $\forall$ means any $\Omega$ satisfying the Friedmann constraint. In our case $\Omega^2\le \frac{1}{2 \xi (1-3\xi)}$.} All points in this case describe vacuum FLRW space-time with negative spatial curvature. This model is known as the Milne universe with the scale factor $a = c_2 ( t + c_1 )$ and Hubble function $H = \frac{1}{t+c_1}$.  Given that $\Omega_H^2=1/2$, one finds that $c_2^2=|k|$. From the definitions of the effective energy and pressure~\eqref{eq:EfEn},~\eqref{eq:EfPr} we get that $\epsilon_e=0$ and $p_e=0$. This implies that we have a vacuum universe dominated by negative curvature and the effective equation of state parameter~\eqref{eq:EoS} is undefined. Furthermore, this implies that $D\rightarrow 0$ for $t\rightarrow\infty$.  Since $\Omega_V=0$ and $\Omega_{\psi}=0$ in the critical point, we necessarily have that $\dot{\psi}\rightarrow 0$ and $V\rightarrow0$, both faster than $D$ approaches zero.  Since we do not have a specific form for the potential, the limiting value of $\Omega_{\partial V}$ remains unspecified. 

 For the line with $\Omega_H=\frac{1}{\sqrt{2}}$, which we call $D_{+}$, we get the eigenvalues 
\begin{equation}
    \{\lambda^{D_{+}}_{i}\}=\{0,0,\frac{1}{\sqrt{2}},-\sqrt{2},-\frac{1}{\sqrt{2}}\},
\end{equation}
and for $\Omega_H=-\frac{1}{\sqrt{2}}$, which we call $D_{-}$ , we get 
\begin{equation}
    \{\lambda^{D_{-}}_{i}\}=\{0,0,-\frac{1}{\sqrt{2}},\frac{1}{\sqrt{2}},\sqrt{2}  \}.
\end{equation}
  The mixed character of the eigenvalues identify these critical planes as saddles.

\section{Specific potential cases} \label{sec:specpot}

Once a form of potential is chosen, the system is completely specified and the variable $\Omega_{\partial V}$ becomes redundant. In the most general case $\Omega_{\partial V}$ has to be a function of $\Omega$ only, because $V=V(\psi)$.   This fact allows to rewrite $\Omega_{\partial V}'$ as
\begin{equation}
 \Omega_{\partial V}' = \frac{\partial \Omega_{\partial V}}{\partial \Omega}\, \Omega'\, .
\end{equation}
Using now eqs.~\eqref{eq:omega} and \eqref{eq:omegadv} we obtain the following differential equation (for $\Omega_{\psi}\neq 0$):
\begin{equation}\label{eq:gendiff}
 \frac{\partial \Omega_{\partial V}}{\partial \Omega}\, \left( 1- \xi\, \Omega^2 \right)^{3/2} = \Omega_{\partial V}^2\, \left(\Gamma(\Omega) -1 \right).
\end{equation}
 To consider $\Omega_{\partial V}=\Omega_{\partial V}(\Omega)$ has been our assumption up to here.  One can however make a different, less general assumption, like in \cite{Hrycyna2010} where it is assumed that $\Omega=\Omega(\Omega_{\partial V})$ and $\Gamma=\Gamma(\Omega_{\partial V})$: this implies that  $\Omega_{\partial V}=\Omega_{\partial V}(\Omega)$ has to be an invertible function, which might not always be the case. In the cases in which this is true, it holds that
\begin{align}
    \psi=\int \frac{d\Omega_{\partial V}}{\Omega_{\partial V}^2(\Gamma(\Omega_{\partial V})-1)} ,
\end{align}
{\it cfr.} eq.~(2.16) in the reference above.

Dropping the  discussion of general potential forms, in this section we are going to focus our study on specific classes of potentials and further restrict our analysis on spatially flat spacetime, which corresponds to an invariant subset of the system. In particular, once the form of potential is chosen, the system reduces to 4 dimensions; further, by assuming $\Omega_H=\pm 1$ and by employing the Friedmann constraint the system effectively reduces to 2-dimensions, evolving on the $(\Omega$, $\Omega_{\psi})$ plane. Thus, the critical points discussed in Sec.~\ref{sec:CritPoi} and singularities of the system (see Sec.~\ref{sec:GenFeat}) will be depicted in the invariant subsets. 

Some of these points are independent of the form of potentials and hence will be present in every specific case that we will discuss below.  These points are
\begin{enumerate}

\item \textit{Big Bang and Big Crunch:} \label{item:bigbang}\\
There are two points $S_+$ on the invariant subsets $\Omega_H=1$ at $\lbrace \Omega,\Omega_\psi  \rbrace\rightarrow\lbrace \pm\sqrt{\frac{1}{2\xi(1-3\xi)}},\mp\sqrt{\frac{6\xi}{1-6\xi}}\rbrace$, which are the singular points of the system and act like Big Bang sources. Moreover, there are two points $S_-$ on the invariant subset $\Omega_H=-1$ at $\lbrace \Omega,\Omega_\psi  \rbrace\rightarrow\lbrace\pm\sqrt{\frac{1}{2\xi(1-3\xi)}},\pm\sqrt{\frac{6\xi}{1-6\xi}}\rbrace$: these are also singular points of the system and act like Big Crunch sinks.  In order to recognize the cosmological character of such points, recall the definition of the evolution parameter of system $\tau = \pm \ln a$, where plus/minus applies to the expanding/collapsing dynamics: as the critical points are approached along the trajectories we have that the parameter $\tau\rightarrow\mp\infty$ and hence in both cases $a\rightarrow0$.
 
 \item  \textit{Radiation-like transient phase:}\label{item:radiation}\\
One can find saddle points $C_+$ or $C_-$ with coordinates $\lbrace \Omega,\Omega_\psi  \rbrace\rightarrow\lbrace \pm \frac{1}{\sqrt{2\xi}},0\rbrace$ in each invariant subset. These points describe a radiation-like universe since $w_{e}=1/3$ (see Table~\ref{tab:CritPoints}), and evolution flows around them define a possible radiation-like transition phase of the universe.
\end{enumerate}
In order to find the locations of the \textit{de-Sitter points} $B_\pm$ in the invariant subsets, one needs instead to specify the form of the potential.

\subsection{$\Gamma=1$: exponential potentials}

\begin{figure*}
    \centering
       { \includegraphics[width=0.48\textwidth]{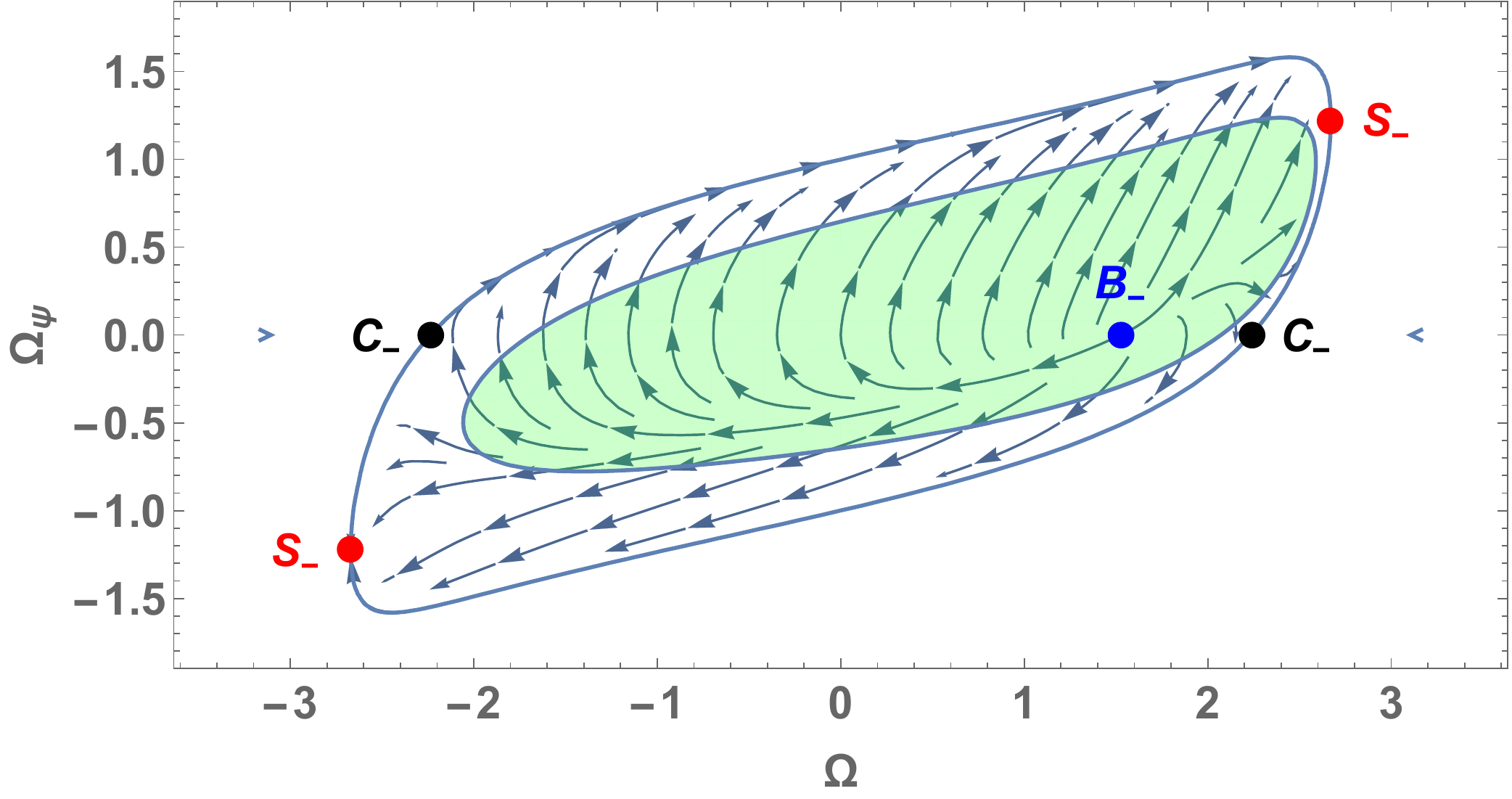}
    \quad
        \includegraphics[width=0.48\textwidth]{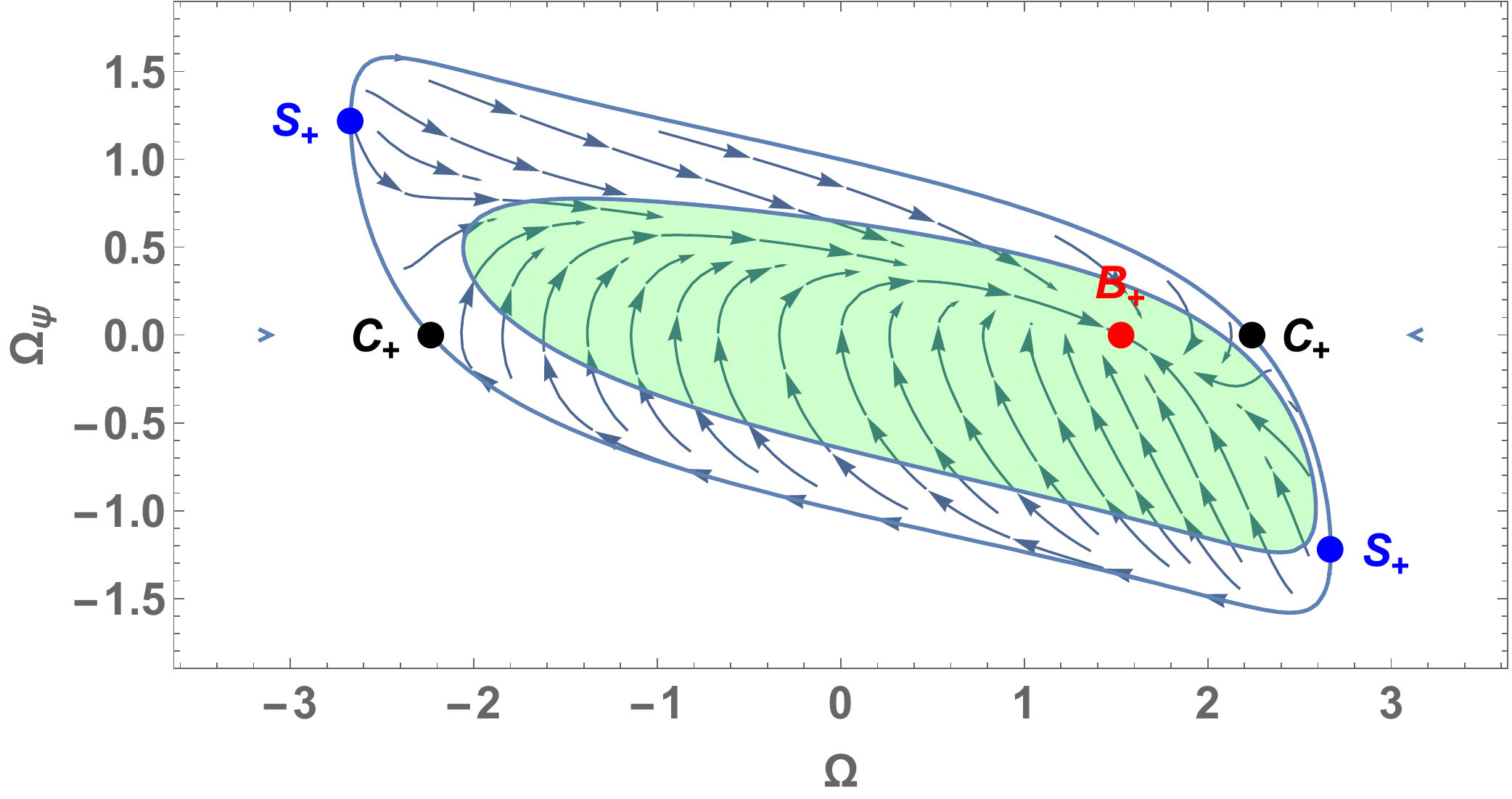}}
      {  \includegraphics[width=0.48\textwidth]{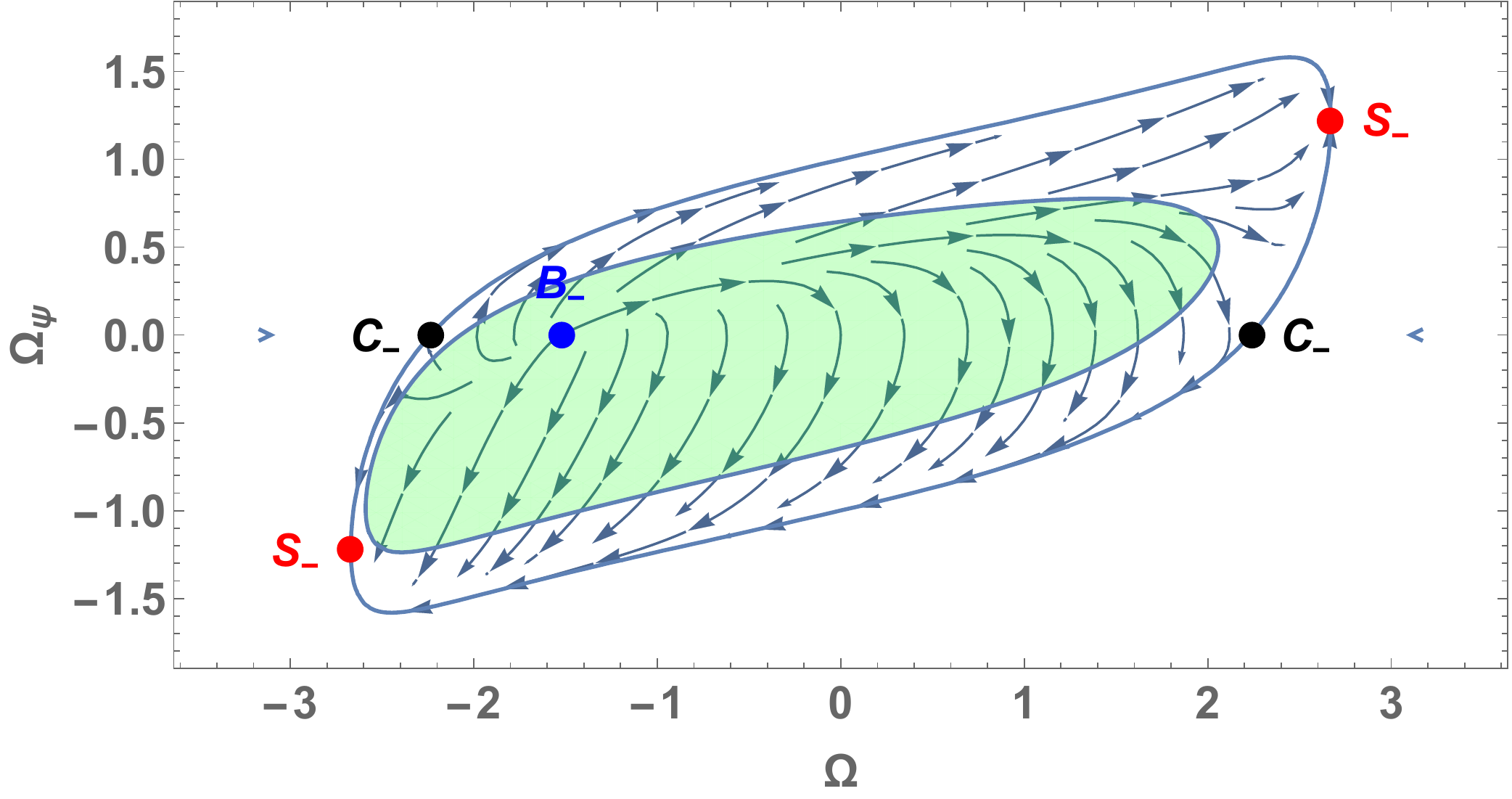}
    \quad
        \includegraphics[width=0.48\textwidth]{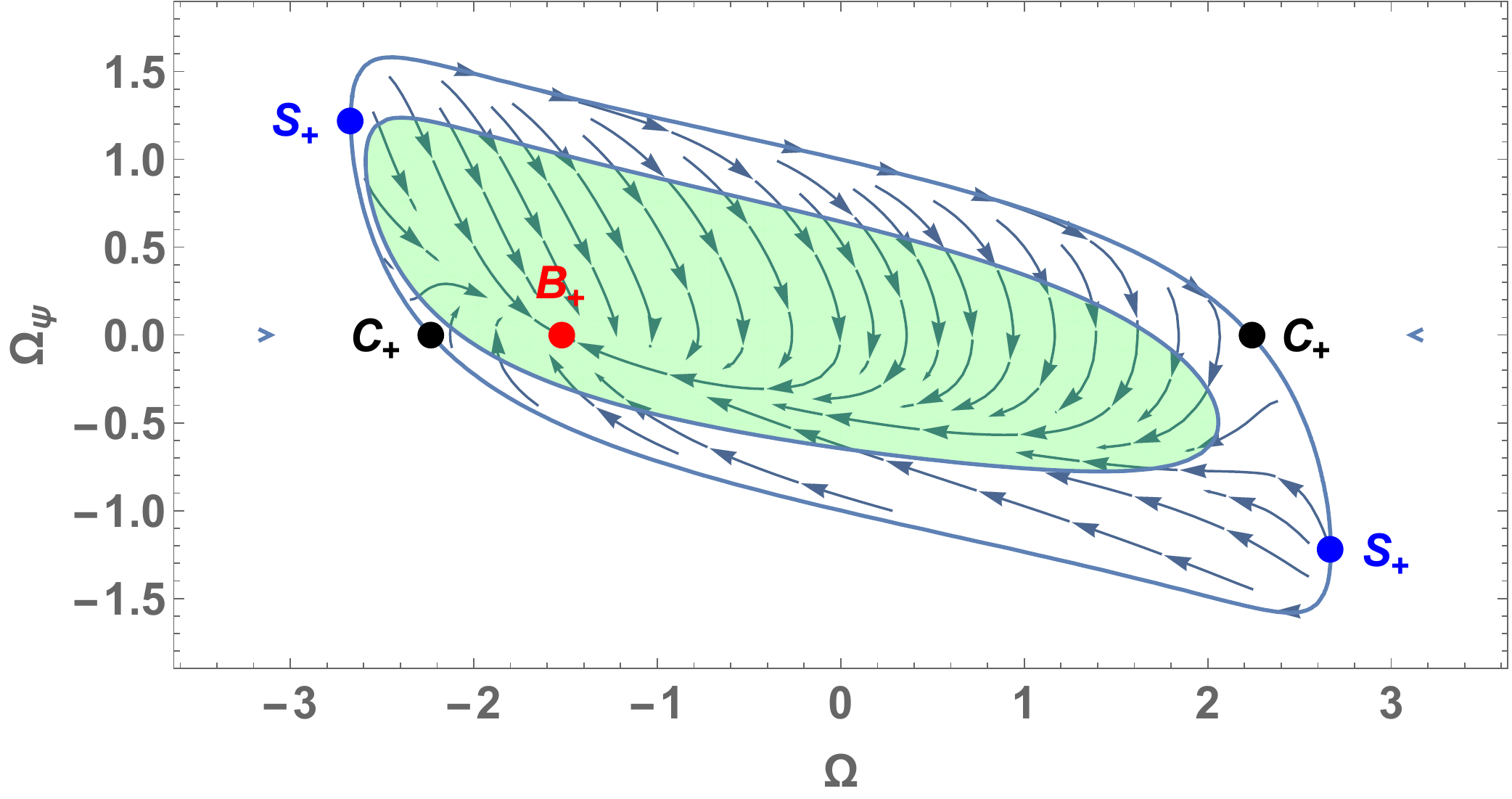}}
      {  \includegraphics[width=0.48\textwidth]{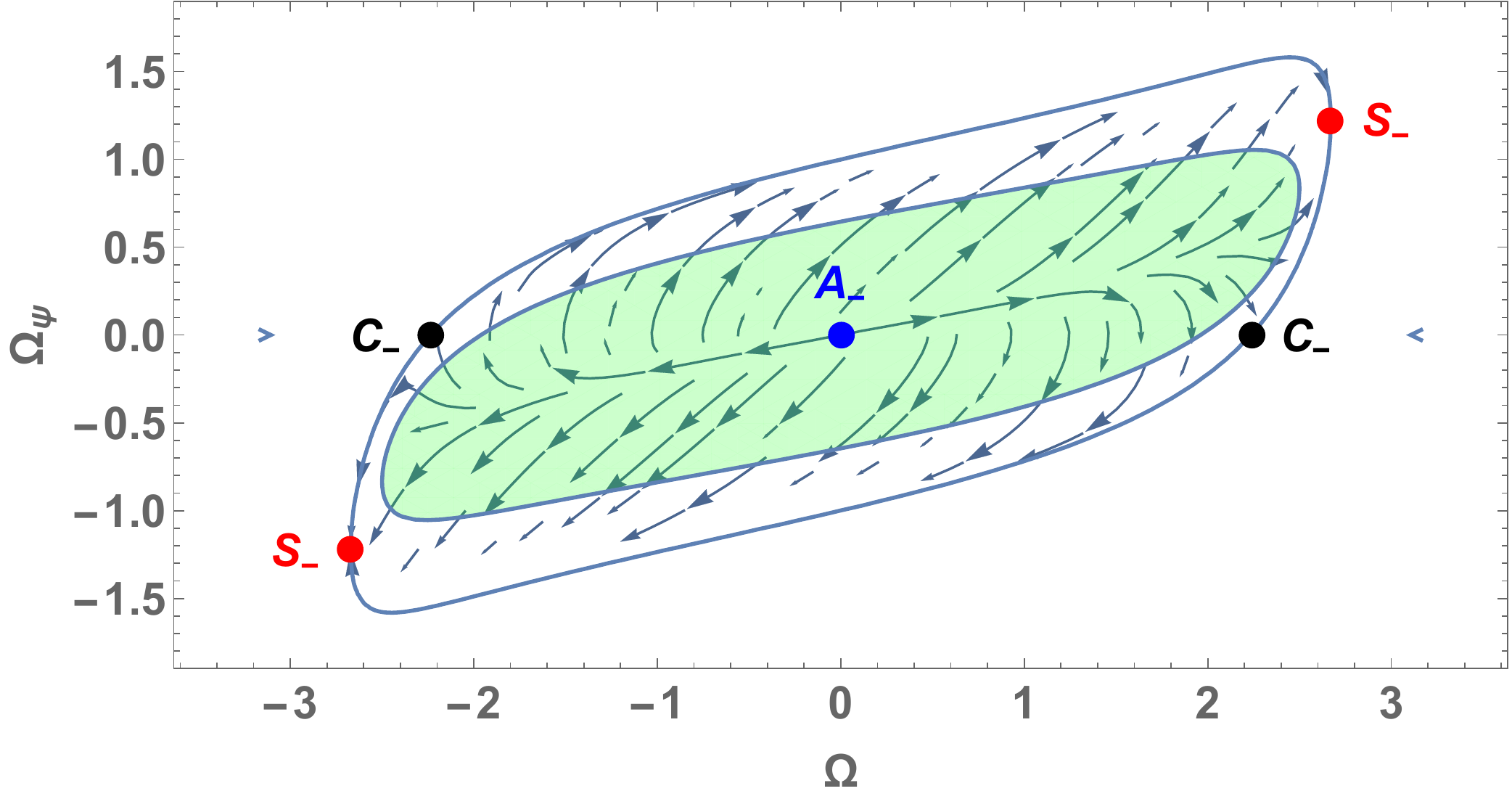}
    \quad
        \includegraphics[width=0.48\textwidth]{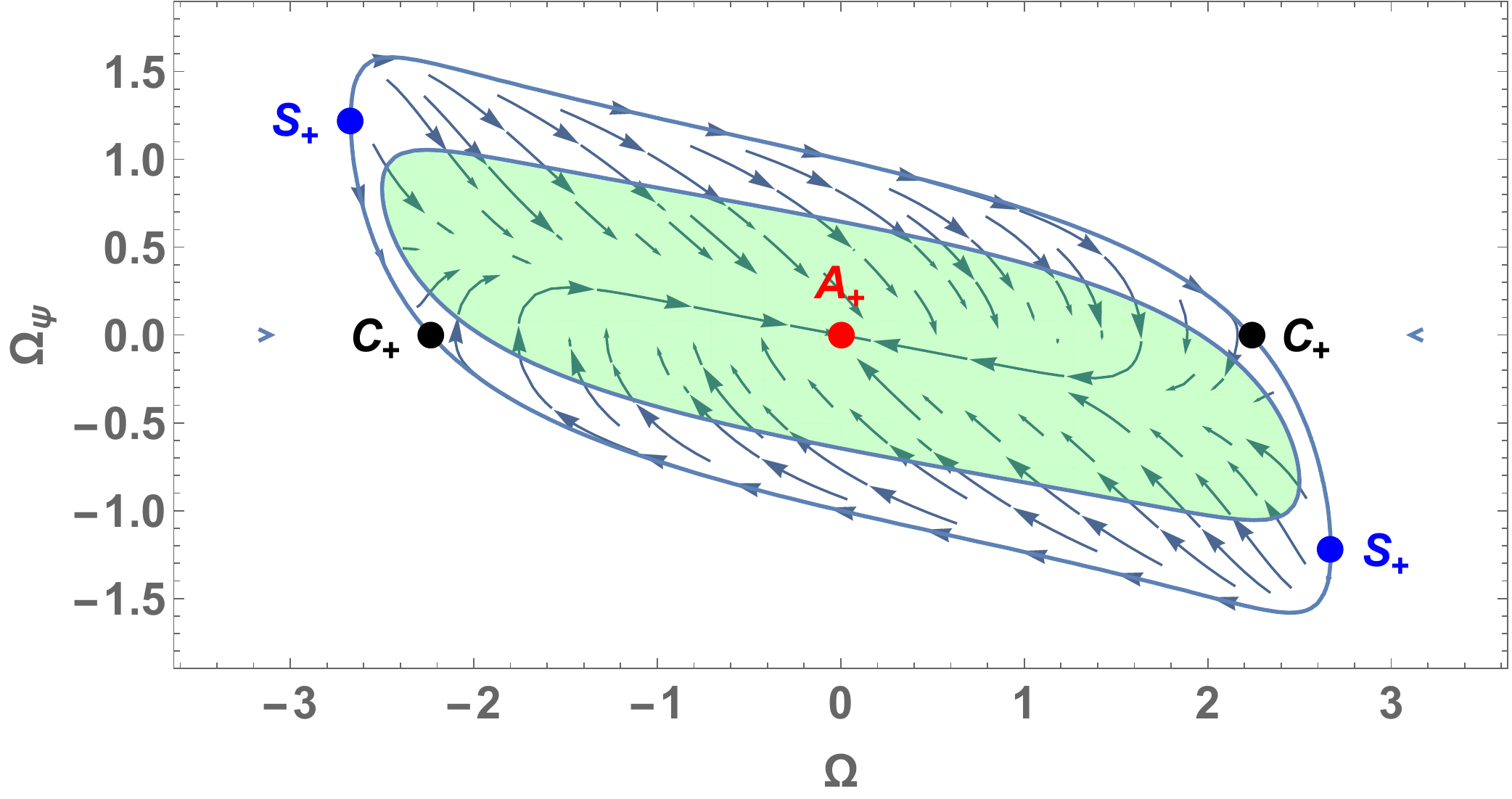}}
    \caption{Some invariant subsets $\Omega_{\partial V}=\textrm{const.}$ for $\xi=\frac{1}{10}$. The left column of panels shows $\Omega_H=-1$ cases, while the right column shows the $\Omega_H=1$ cases. Upper panels show $\Omega_{\partial V}=-1$, the middle ones $\Omega_{\partial V}=1$ and the Lower ones $\Omega_{\partial V}=0$.  Blue dots identify sources, red dots are sinks and black ones are saddle points.  The  green areas denote the phase of accelerated expansion $q<0$ .} \label{fig:FlatSubset}
\end{figure*}

\begin{figure}[ht]
    \centering
      \includegraphics[width=0.5\textwidth]{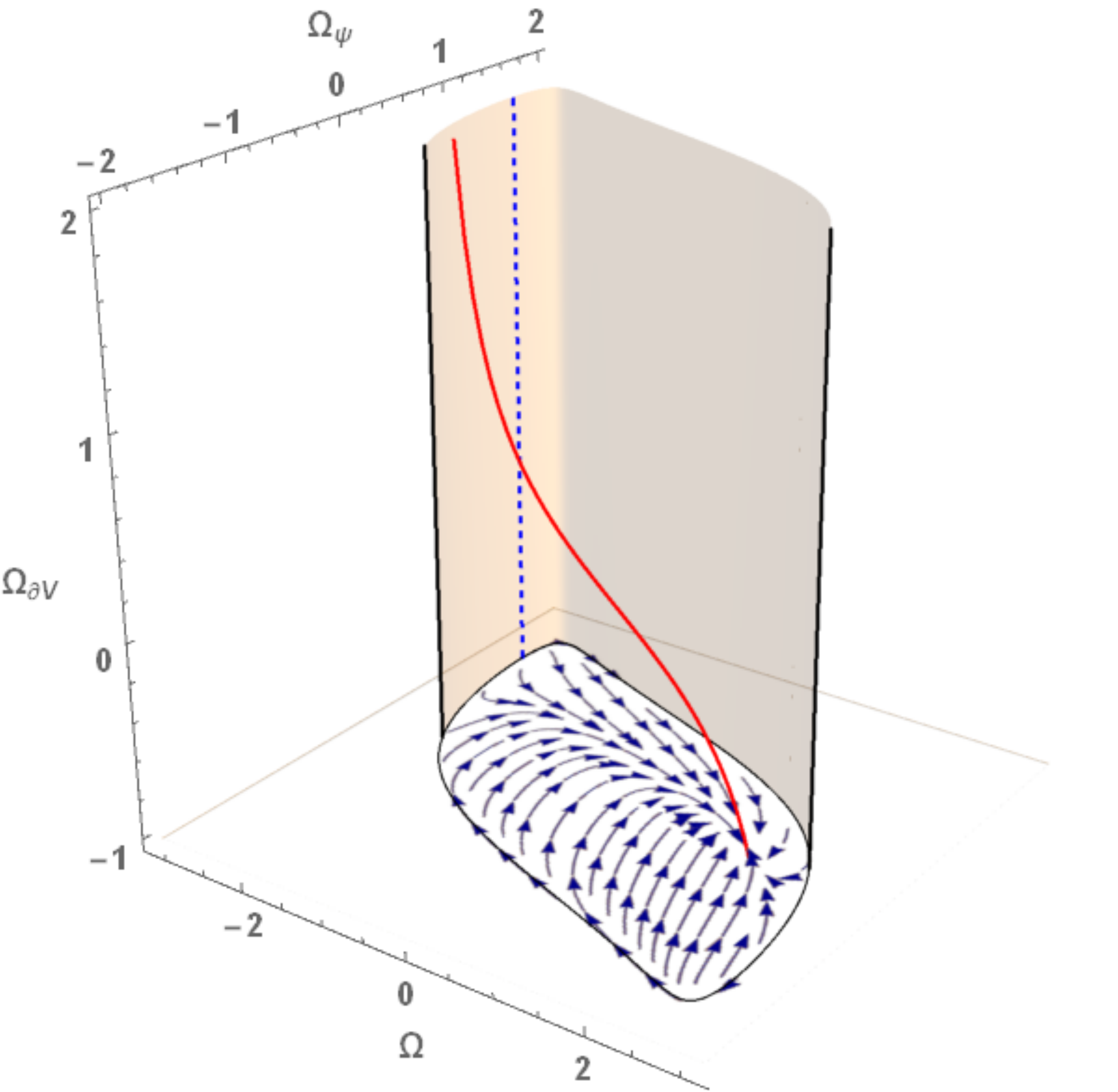}
    \caption{Global view of the parameter space for the system with $\Gamma=1$, $\Omega_H=1$ and $\xi=1/10$.  The dynamics takes place inside every horizontal plane with constant $\Omega_{\partial V}$; the case $\Omega_{\partial V}=-1$ is shown as representative.  For a better view, the closed boundary surface corresponding to $\Omega_V = 0$ is cut along the plane $\Omega_\psi=0$.  The blue dashed line is one set of sources, the black lines are the saddles and the red line is the set of future attractors.} \label{fig:exp_flat}
\end{figure}
For the special cases when $\Gamma=1$ it holds that $\Omega_{\partial V}=\textrm{const.}$,
thus in this case the potentials are of the form
$$V=V_0 e^{\Omega_{\partial V} \psi}.$$
Each system with $\Omega_H=1$ has one sink at the coordinate:
\begin{itemize}
    \item $\lbrace \Omega,\Omega_\psi  \rbrace\rightarrow\lbrace \sqrt{\frac{1}{2\xi}-\frac{1}{\sqrt{\Omega_{\partial V}^2\xi+4\xi^2}}},0\rbrace$  for $\Omega_{\partial V}<0$,
    \item $\lbrace \Omega,\Omega_\psi  \rbrace\rightarrow\lbrace -\sqrt{\frac{1}{2\xi}-\frac{1}{\sqrt{\Omega_{\partial V}^2\xi+4\xi^2}}},0\rbrace$ for $\Omega_{\partial V}>0$,
    \item $\lbrace \Omega,\Omega_\psi  \rbrace\rightarrow \lbrace 0,0 \rbrace$ for $\Omega_{\partial V}=0$.
\end{itemize}
These three cases correspond to the points $B_+$ depicted on the right column of Fig.~\ref{fig:FlatSubset}, from top to bottom.  With $\Omega_H=-1$ instead the system has a source at the coordinate:
  \begin{itemize}
      \item $\lbrace \Omega,\Omega_\psi  \rbrace\rightarrow\lbrace \sqrt{\frac{1}{2\xi}-\frac{1}{\sqrt{\Omega_{\partial V}^2\xi+4\xi^2}}},0\rbrace$ for $\Omega_{\partial V}<0$,
      \item $\lbrace \Omega,\Omega_\psi  \rbrace\rightarrow\lbrace -\sqrt{\frac{1}{2\xi}-\frac{1}{\sqrt{\Omega_{\partial V}^2\xi+4\xi^2}}},0\rbrace$ for $\Omega_{\partial V}>0$,
      \item $\lbrace \Omega,\Omega_\psi  \rbrace\rightarrow \lbrace 0,0 \rbrace$ for $\Omega_{\partial V}=0$.
  \end{itemize}
These cases correspond to the points $B_-$ depicted in the left column of Fig.~\ref{fig:FlatSubset}, from top to bottom.  For a global view see Fig.~\ref{fig:exp_flat}: horizontal slicings correspond to different constant values of $\Omega_{\partial V}$.

\subsection{Constant $\Gamma\neq1$}

\begin{figure}[ht]
    \centering
      \includegraphics[width=0.5\textwidth]{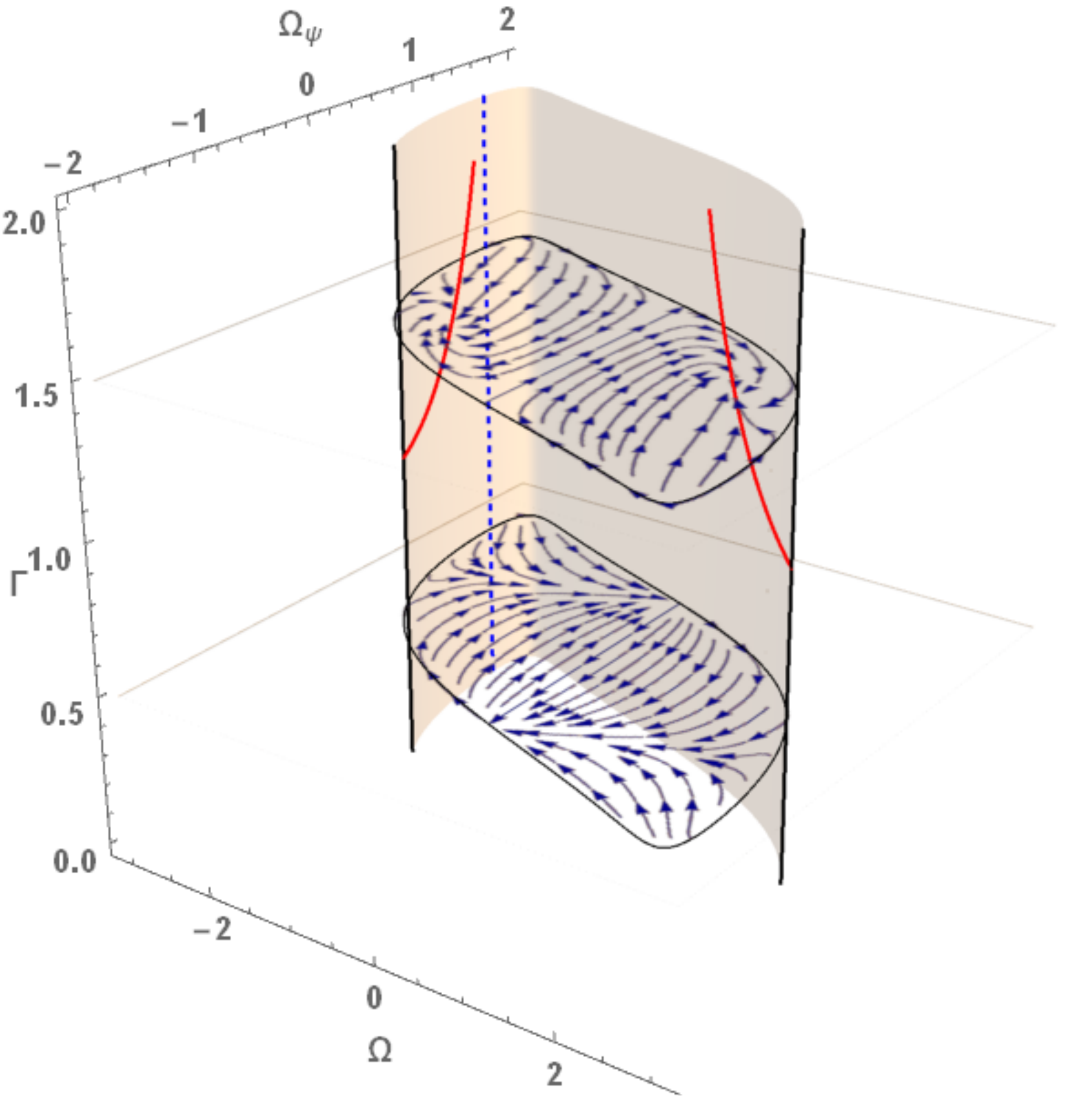}
    \caption{Global view of the parameter space for the system with $\Gamma\neq1$ and constant, $\Omega_H=1$, $\xi=1/10$ and $c_1=0$.  The dynamics takes place inside every horizontal plane with constant $\Gamma$; the cases $\Gamma=1/2$ and $\Gamma=3/2$ are shown as representative.  For a better view, the closed boundary surface corresponding to $\Omega_V = 0$ is cut along the plane $\Omega_\psi=0$.  The blue dashed line is the set of sources, the black lines are the saddles and the red lines are the sets of future attractors.} \label{fig:gammaconst_flat}
\end{figure}

Assuming $\Gamma\neq1$ and constant we can integrate Eq.~\eqref{eq:gendiff} and obtain
\begin{equation}\label{eq:fun_omegadv2}
 \Omega_{\partial V} = \frac{\sqrt{1-\xi\,  \Omega ^2}}{(1-\Gamma ) \Omega - c_1\, \sqrt{1-\xi\,  \Omega ^2}}\, .
\end{equation}
Using the definition of $\Omega_{\partial V}$, this expression can be integrated again, resulting in the following family of potentials:
\begin{equation}\label{eq:fun_V}
 V = V_0\, \Big( (1-\Gamma ) \psi - c_1 \Big)^{\frac{1}{1-\Gamma }}
\end{equation}
The denominator of Eq.~\eqref{eq:fun_omegadv2} introduces a singular line when
\begin{align}\label{eq:singC1}
    c_1-\frac{(1-\Gamma) \Omega}{\sqrt{1-\xi\, \Omega^2}}\,=\,0\,  
\end{align}
and the location of the singular line is at
\begin{align}\label{eq:singOm}
    \Omega_s^2=\frac{c_1^2}{(1-\Gamma)^2+\xi\, c_1^2}\, <\frac{1}{2\xi\, (1-3\,\xi)} 
\end{align}
for $\forall\, \Omega_\psi$ inside the Friedmann constraint apart the Friedmann constraint's outer edge ($\Omega_V=0$). The inequality in Eq.~\eqref{eq:singOm} comes from the restrictions of $\Omega_s$ between the singularities~\eqref{eq:sing} discussed in Sec.~\ref{sec:GenFeat}, also shown in Figs.~\ref{fig:FlatSubset},~\ref{fig:exp_flat}. Further, Eq.~\eqref{eq:singC1} indicates that:
\begin{itemize}
 \item when $\Gamma<1$ then the sign of $\Omega$ (Eq.~\eqref{eq:singOm}) has to be the same as of $c_1$,
 \item while when $\Gamma>1$, then the sign of $\Omega$ (Eq.~\eqref{eq:singOm}) has to be the opposite of $c_1$.
\end{itemize}

\begin{figure}[ht]
    \centering
       {\includegraphics[width=0.45\textwidth]{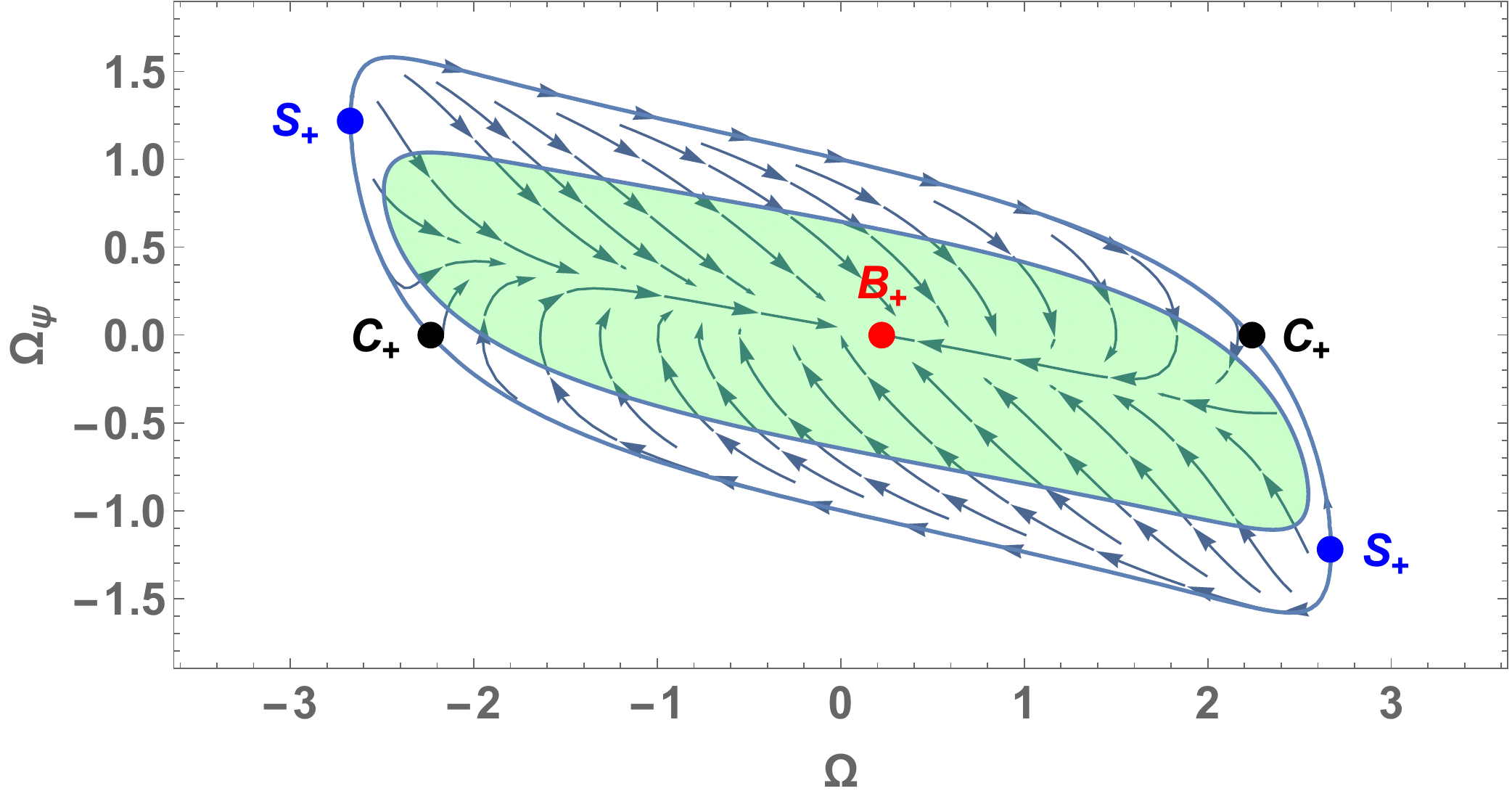}
        \includegraphics[width=0.45\textwidth]{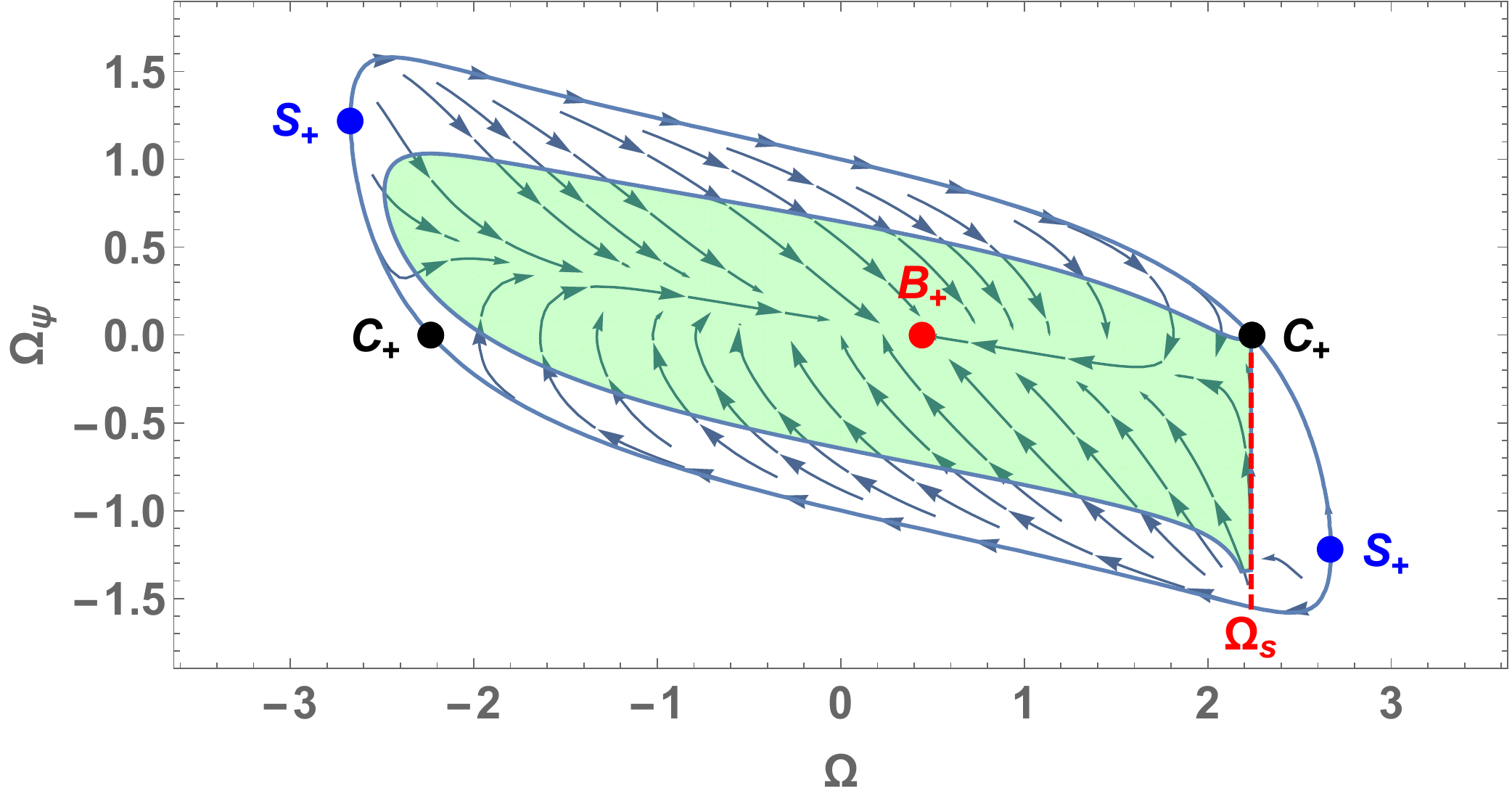}
        \includegraphics[width=0.45\textwidth]{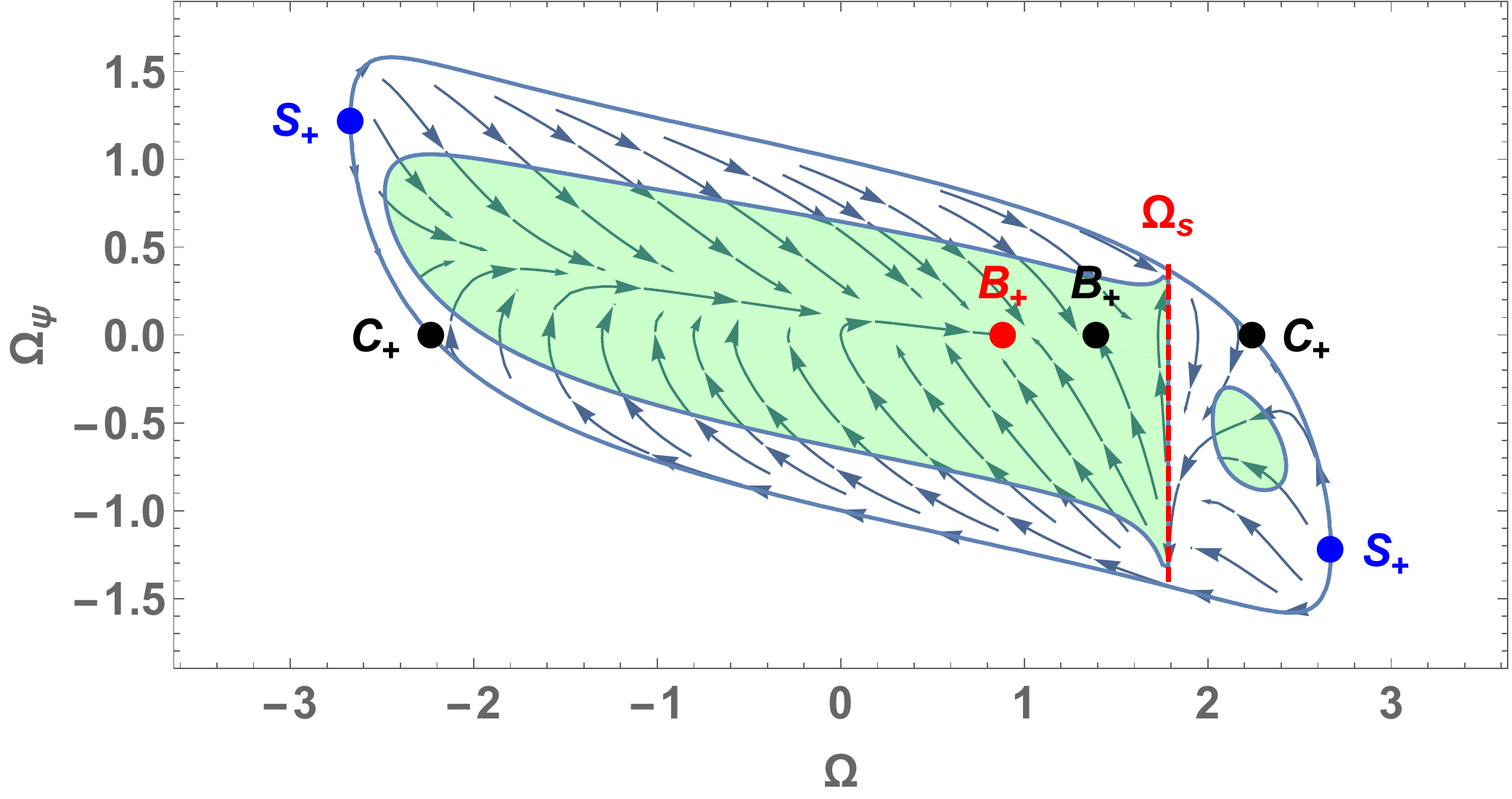}}
    \caption{For the case $\xi=\frac{1}{10}$ invariant subsets for $\Gamma=-1$. The upper panel shows $c_1=\sqrt{\frac{(1-\Gamma)^2}{\xi}}+5$, the middle panel $c_1=\sqrt{\frac{(1-\Gamma)^2}{\xi}}$ and the bottom panel $c_1=\sqrt{\frac{(1-\Gamma)^2}{\xi}}-2$ . The  green areas denote the phase of accelerated expansion $q<0$ .} \label{fig:Gammaless}
\end{figure}

In order to find the critical points $B_{\pm}$ for the potential~\eqref{eq:fun_V} we equate Eq.~\eqref{eq:fun_omegadv2} with the value of $\Omega_{\partial V}$ (Eq.~\eqref{eq:genOmegdv}) presented in Sec.~\ref{sec:CPB}. Note that by doing this we select a particular case of the general treatment. This provides the following relation that will be helpful in order to determine which critical points are inside the Friedmann constraint:
\begin{equation}\label{eq:c1}
  c_1 \Omega\, = \frac{1+2\, \xi\, \Omega^2\, \left(1-2\Gamma\right)}{4\, \xi\, \sqrt{1-\xi\, \Omega^2}}
\end{equation}
Solving the equation above provides four solutions for $\Omega$:
\begin{align}\label{eq:roots}
  \Omega_{\pm,\pm}=\pm \frac{1}{\sqrt{2 \xi}}\sqrt{\frac{(4 c_1^2 \xi +2\Gamma-1) \pm 2  \sqrt{c_1^2 \xi (4 c_1^2 \xi + 4 \Gamma-3)}}{4(c_1^2 \xi+\Gamma(\Gamma-1))+1}}\, ,
\end{align}
where the first set of signs in the definition refers to the global sign, while the second one refers to the sign under square-root. \footnote{One can easily check that these 4 roots appear as critical points of the general dynamical system when $\Omega_{\partial V}$ is given by Eq.\eqref{eq:fun_omegadv2}.}  The existence of such critical points depends on the relative sign of $c_1$ and $\Omega$ as expressed in Eq.~\eqref{eq:c1}; then, if a root exists, we need additionally a condition for it to satisfy the Friedmann constraint.  In Tables \ref{tab:gammal12}--\ref{tab:gammag1} we give the exact ranges of parameters in which the roots~\eqref{eq:roots} exist.  In particular:
\begin{enumerate}
 \item if $\Gamma\leq1/2$, then the numerator of Eq.~\eqref{eq:c1} is positive and hence $c_1$ and $\Omega$ should have the same sign, {\it i.e.} $c_1\cdot \Omega>0$.  This implies that only the two roots among those in Eq.~\eqref{eq:roots} with the same global sign as $c_1$ will be allowed.  Additionally, the Friedmann constraint and the combination of parameters will define whether these two roots will appear or not, as shown in Table \ref{tab:gammal12}.  When both roots exist, they appear on the same side of the singular line $\Omega_s$: if $c_1>0$, then $\Omega_{+-}$ is a sink (source) while $\Omega_{++}$ is a saddle; if $c_1<0$, then $\Omega_{--}$ is a sink (source) while $\Omega_{-+}$ is a saddle.
 \item if $1/2<\Gamma<1$, then  
 \begin{equation}
  1+2\, \xi\, \Omega^2\, (1-2\, \Gamma) > 1-2\, \xi\, \Omega^2 > 0\, ,
 \end{equation}
 where the last inequality comes from the Friedmann constraint.  This has the same implication as the case above about the relative signs of $c_1$ and $\Omega$. In this case there is at most one root, see Table~\ref{tab:gammabetween} for the details.
 \item if $\Gamma>1$, the numerator of Eq.~\eqref{eq:c1} has two roots $\Omega = \pm \left( 2\, \xi\, |1-2\Gamma| \right)^{-1/2}$: between these roots the numerator is positive, while outside it is negative.  The sign of $c_1\cdot \Omega$ has to be the same as the one of the numerator, thus determining which of the roots in Eq.~\eqref{eq:roots} are present, see Table~\ref{tab:gammag1}.  When both roots exist, the singular line $\Omega_s$ lies between them; in this case $B_+$ ($B_-$) retain their sink (source) nature.
\end{enumerate}
It is worth stressing that the critical points $\Omega_{\pm\pm}$, denoted as $B_{\pm}$ in Figs.\ref{fig:Gammaless}-\ref{fig:gamma3o4}, move inside the Friedmann constraint along the line $\Omega=0$ when the parameters change in the ranges allowed by Tables \ref{tab:gammal12},\ref{tab:gammabetween} and \ref{tab:gammag1}. In cases 1. and 3. above, when $\xi\, c_1^2 \rightarrow (1-\Gamma)^2$, then $\Omega_{\pm+}$ approaches the position of the radiation-like saddle points $C_{\pm}$; in case 2., the same happens for $\Omega_{\pm-}$.  However, the cosmological interpretation of points $B_{\pm}$ is preserved as they move and Table 1 excludes the case $\Omega^2=1/2\xi$ for the de Sitter sinks/sources: hence the de Sitter character which is preserved as the points move is not in contradiction with the radiation character on the boundary in the above-mentioned limit.

In Fig.~\ref{fig:Gammaless} we show the case $\Gamma=-1$, corresponding to a potential $V\propto\sqrt{\psi}$ which has some interesting dynamical property but is otherwise physically questionable.  From top to bottom, we change gradually the parameter $c_1$ in order to show how one of the de Sitter points $B_+$ appears inside the Friedmann constraint and changes its character from sink to saddle.  In the top panel such point is outside the constraint; in the middle panel, it coincides with the radiation-like saddle $C_+$; and in the bottom panel, it appears as a saddle on the right-hand side of the singular segment.  The dynamical setup of the bottom panel is quite intriguing, as it presents de Sitter phases both as a transient and as an asymptotic attractor; note however that the potential is complex on the left-hand side of the singular segment, so one cannot give a physical interpretation to such dynamics.  In the next subsection instead we will present some physically meaningful cases.

\begin{table}[]
    \centering
    \begin{tabular}{c||c | c | c}
         & $c_1<0$ & $c_1 = 0$ & $c_1>0$ \\
         \hline
        $\Omega_{++}$ & -- & -- & $\frac{3}{4}-\Gamma \leq \xi\, c_1^2 < (1-\Gamma)^2$ \\ 
        $\Omega_{+-}$ & -- & -- & $\xi\, c_1^2 > \frac{3}{4}-\Gamma$\\
        $\Omega_{-+}$ & $\frac{3}{4}-\Gamma \leq \xi\, c_1^2 < (1-\Gamma)^2$ & -- & -- \\
        $\Omega_{--}$ & $\xi\, c_1^2 > \frac{3}{4}-\Gamma$ & -- & --
    \end{tabular}
    \caption{Existence of the roots Eq.~\eqref{eq:roots} for $\Gamma\leq1/2$.}
    \label{tab:gammal12}
\end{table}

\begin{table}[]
    \centering
    \begin{tabular}{c||c | c | c}
         & $c_1<0$ & $c_1 = 0$ & $c_1>0$ \\
         \hline
        $\Omega_{++}$ & -- & -- & -- \\ 
        $\Omega_{+-}$ & -- & -- & $\xi\, c_1^2 > (1-\Gamma)^2$\\
        $\Omega_{-+}$ & -- & -- & -- \\
        $\Omega_{--}$ & $\xi\, c_1^2 > (1-\Gamma)^2$ & -- & --
    \end{tabular}
    \caption{Existence of the roots Eq.~\eqref{eq:roots} for $1/2<\Gamma<1$.}
    \label{tab:gammabetween}
\end{table}

\begin{table}[]
    \centering
    \begin{tabular}{c||c | c | c}
         & $c_1<0$ & $c_1 = 0$ & $c_1>0$ \\
         \hline
        $\Omega_{++}$ & $0<\xi\, c_1^2 < (1-\Gamma)^2$ & $\forall \xi$ & -- \\ 
        $\Omega_{+-}$ & -- & $\forall \xi$ & $\forall \xi$ \\
        $\Omega_{-+}$ & -- & $\forall \xi$ & $0<\xi\, c_1^2 < (1-\Gamma)^2$ \\
        $\Omega_{--}$ & $\forall \xi$ & $\forall \xi$ & --
    \end{tabular}
    \caption{Existence of the roots Eq.~\eqref{eq:roots} for $\Gamma>1$.}
    \label{tab:gammag1}
\end{table}

\subsection{Physical interpretation}\label{sec:PhysInter}

In this section we focus on the cases with $\Gamma\neq1$ that could have physical interest.  Our discussion will be constrained to $\Omega_H=1$ and we will keep $\xi=1/10$ for our examples.

First of all, our initial requirement of positivity of the potential translates into the choice $\Gamma=\frac{2n+1}{2n}$ for $n\in \mathbb{Z}$: the set of potentials characterized by such exponents are positively defined on the real axis.  Depending on the sign of the integer $n$ one can identify the following classes:
\begin{enumerate}
    \item if $n\in \mathbb{N}^-$, then $\Gamma>1$ and the potentials have the so-called {\it runaway} form:
    \begin{equation}
     V = V_0\, \left( \frac{\psi}{2|n|} + c_1 \right)^{-2 |n|}
    \end{equation}
    \item if $n\in \mathbb{N}^+$, then $1/2\leq\Gamma<1$ and the potentials are positive even powers of the (shifted) field:
    \begin{equation}\label{eq:evenpot}
     V = V_0\, \left( \frac{\psi}{2n} - c_1 \right)^{2n}
      \end{equation}
\end{enumerate}
As one usually considers only potentials with even powers of the field, the cases $\Gamma<1/2$ are excluded.  Note that potentials Eq.~\eqref{eq:evenpot} can be considered as truncated Taylor expansions of more general potentials.

One can check whether a potential defines a mass for the scalar field by analysing the second derivative $\partial^2_{\psi} V$ in a local minimum $\bar{\psi}$ of $V$ itself.  While in the class 1.~there is no such minimum for finite values of the field, in the class 2.~we can distinguish
\begin{itemize}
    \item $n=1\, \Rightarrow\, \Gamma=1/2$ and $(\partial^2_{\psi} V)_{\bar{\psi}} = V_0/2 >0$, massive scalar field;\\
    \item $n\geq2\, \Rightarrow\, (\partial^2_{\psi} V)_{\bar{\psi}} = 0$, massless scalar field.
\end{itemize}
Hence, in the class of positive definite potentials, only the ones with $\Gamma=1/2$ have non-zero mass.  This case corresponds to the simple quadratic potential.

\begin{figure*}[ht]
    \centering
       { \includegraphics[width=0.45\textwidth]{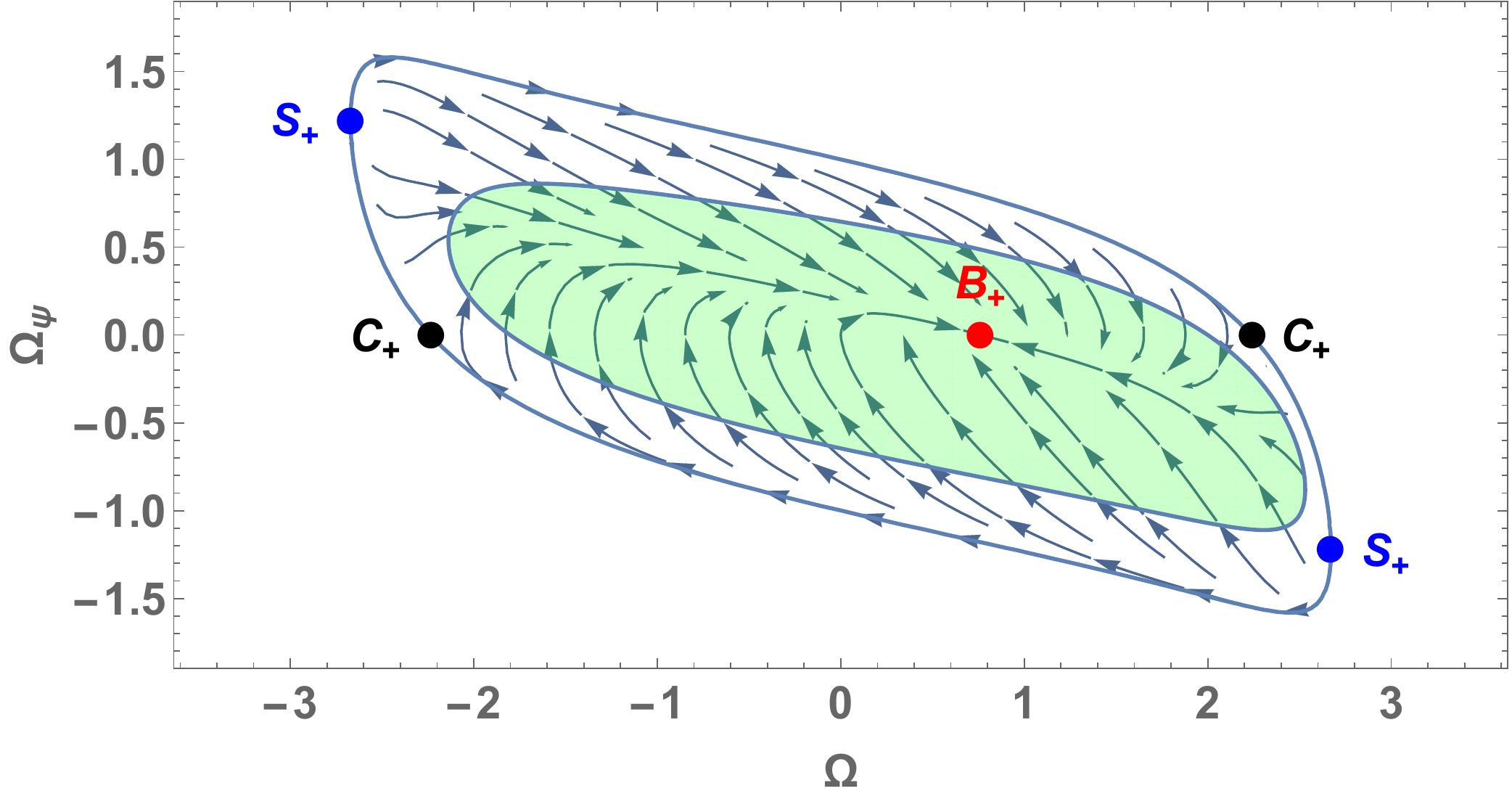}\quad
        \includegraphics[width=0.45\textwidth]{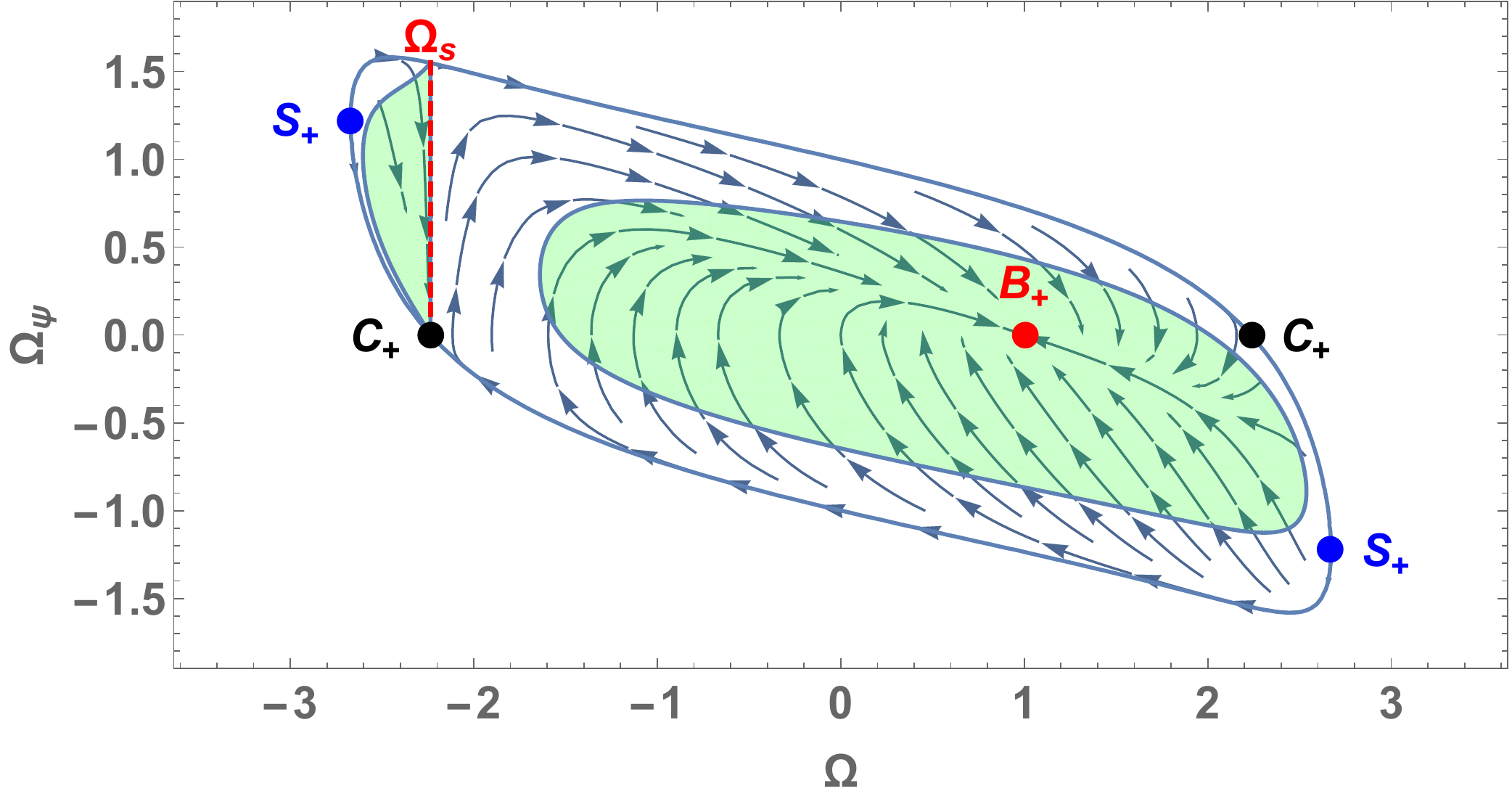}
        \includegraphics[width=0.45\textwidth]{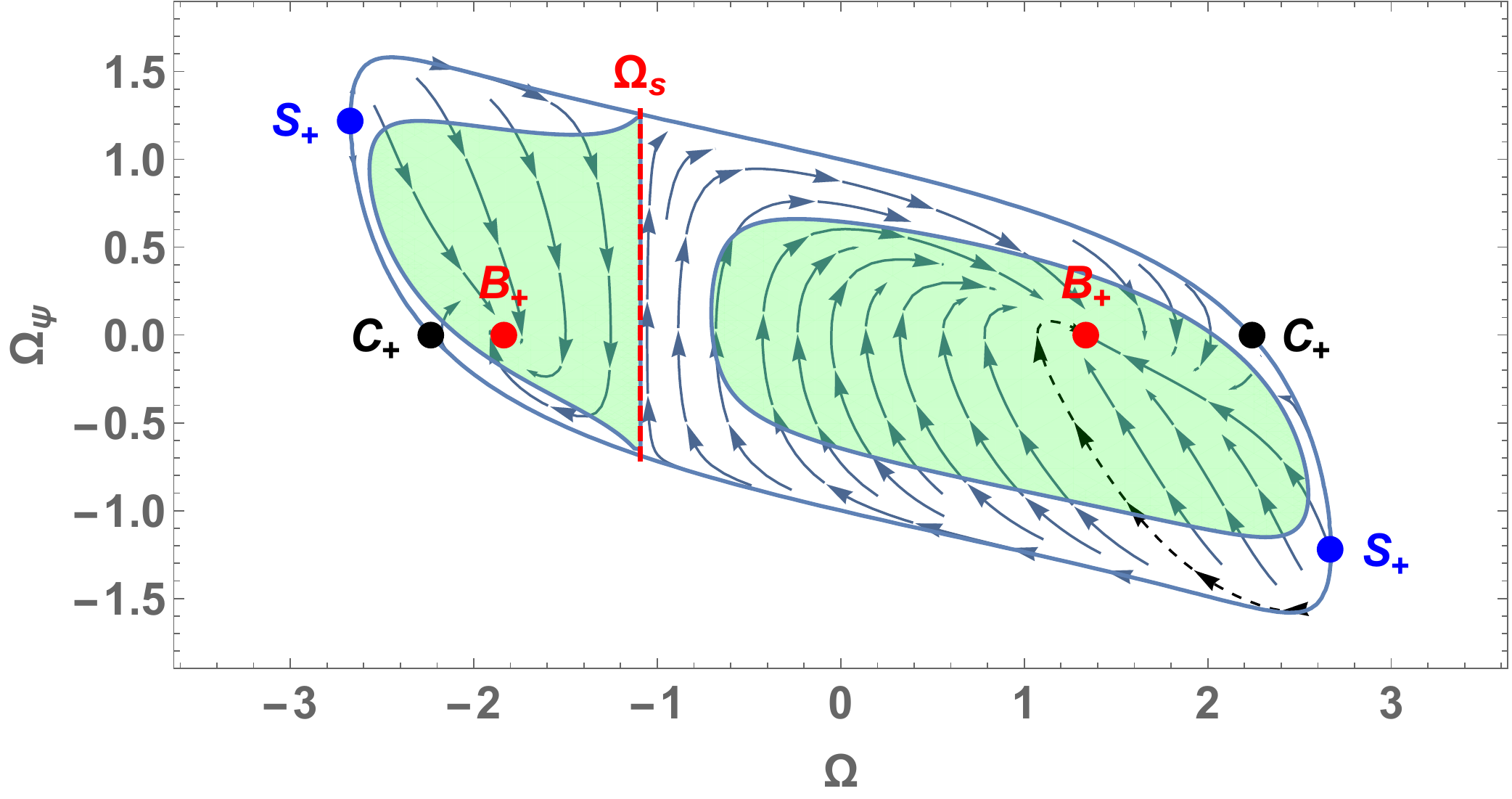}\quad
        \includegraphics[width=0.4\textwidth]{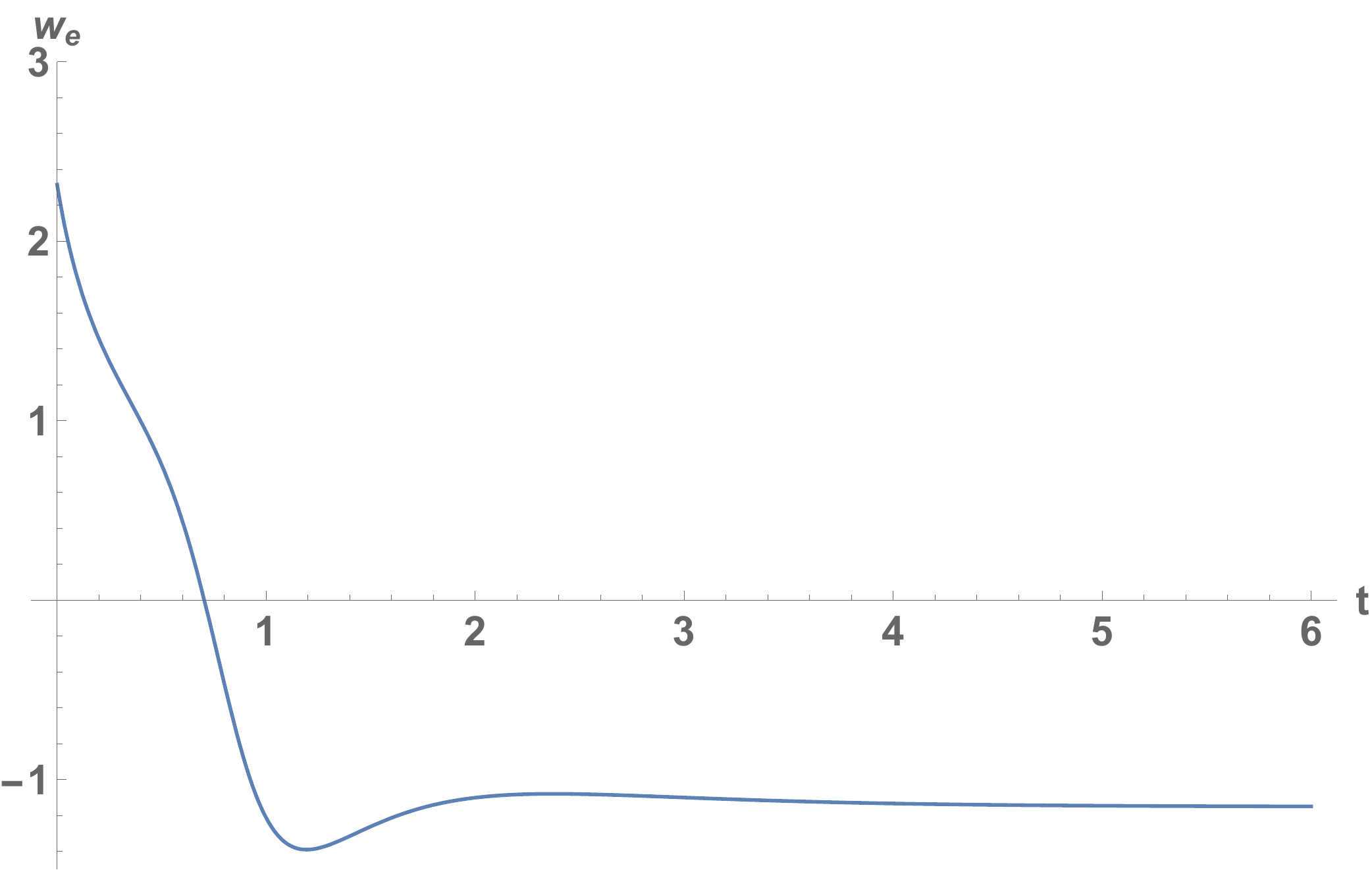}}
       \caption{For the case $\xi=\frac{1}{10}$ invariant subsets for $\Gamma = 3/2$. The upper left panel shows $c_1=\sqrt{\frac{(1-\Gamma)^2}{\xi}}+1$, the upper right panel $c_1=\sqrt{\frac{(1-\Gamma)^2}{\xi}}$ and the bottom ones $c_1=\sqrt{\frac{(1-\Gamma)^2}{\xi}}-1$ . The  green areas denote the phase of accelerated expansion $q<0$. The bottom right panel shows the effective EOS parameter corresponding to the black-dashed trajectory in the left bottom panel, with initial conditions given by $\{ \Omega=5/2 , \Omega_\psi = -1-1/\sqrt{3} \}$.}\label{fig:Gammabig}
\end{figure*}

On the other hand, one could relax the requirement of positivity and well-definedness of the potential on the whole real axis and accept also potentials which are defined only for some ranges of $\psi$.  Potentials of class 1.~diverge in $\psi_0=2 |n| c_1$ and the field is expected to roll down the slope only on one side of $\psi_0$: this translates into a dynamics which is confined only on one side of the singular line \eqref{eq:singOm} in the parameter space.  Hence, in the case of runaway potentials one could in principle allow for any real value in the range $\Gamma>1$ and be careful to consider only the dynamics in the appropriate side of the parameter space.  For instance, in Fig.~\ref{fig:Gammabig} we represent the case $\Gamma=3/2$, for which the runaway-type potential is real and positively defined on the whole real axis except for the singular point $\psi_0$: hence the field can in principle roll down on both sides of the singularity, depending on the initial conditions, and both sides of the parameter space are physically admissible.  For $c_1<\sqrt{(1-\Gamma)^2/\xi}$, the model evolves towards an asymptotic de Sitter attractor on both sides.

\begin{figure}[ht]
    \centering
        {\includegraphics[width=0.45\textwidth]{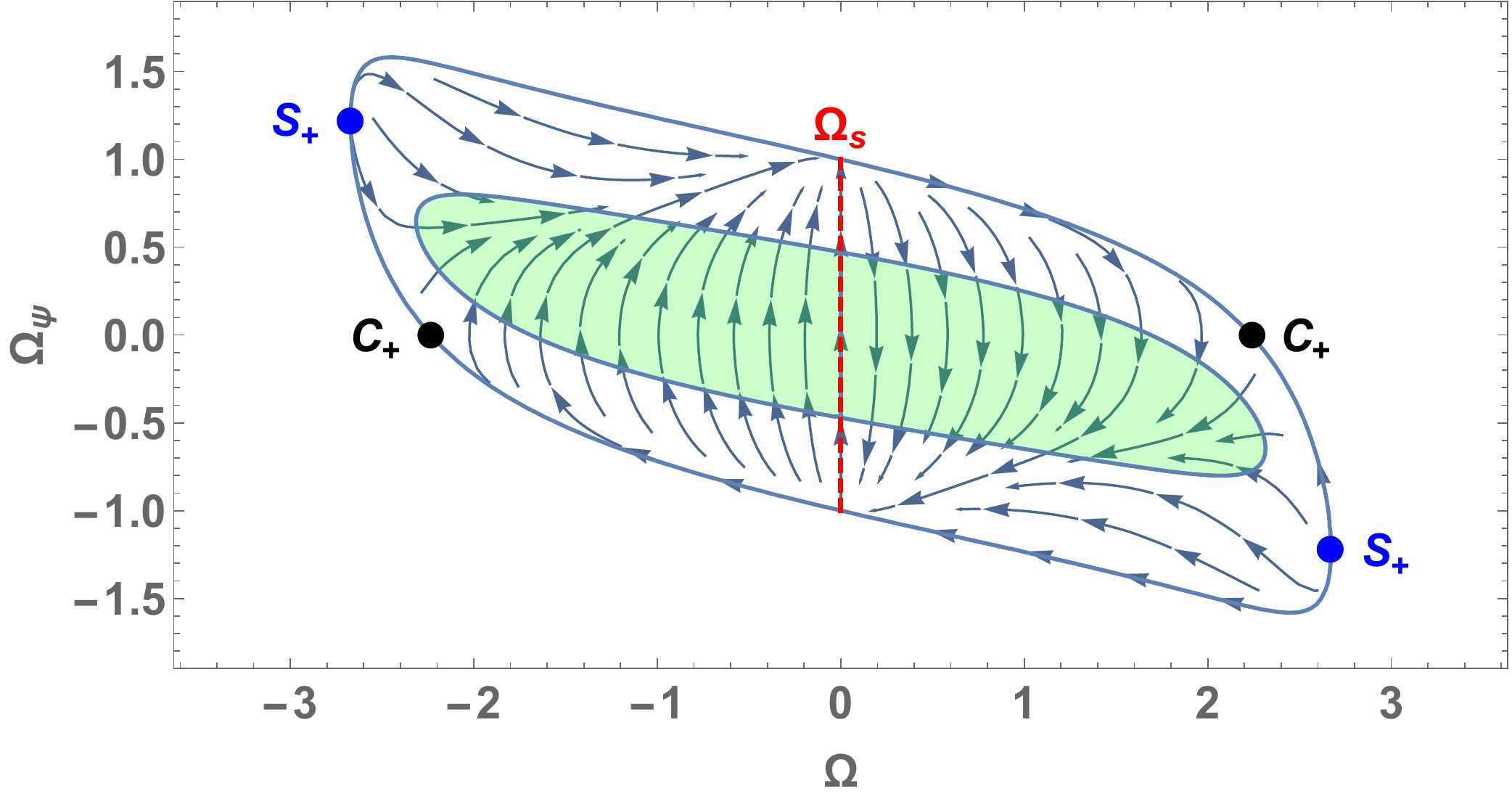}
        \includegraphics[width=0.45\textwidth]{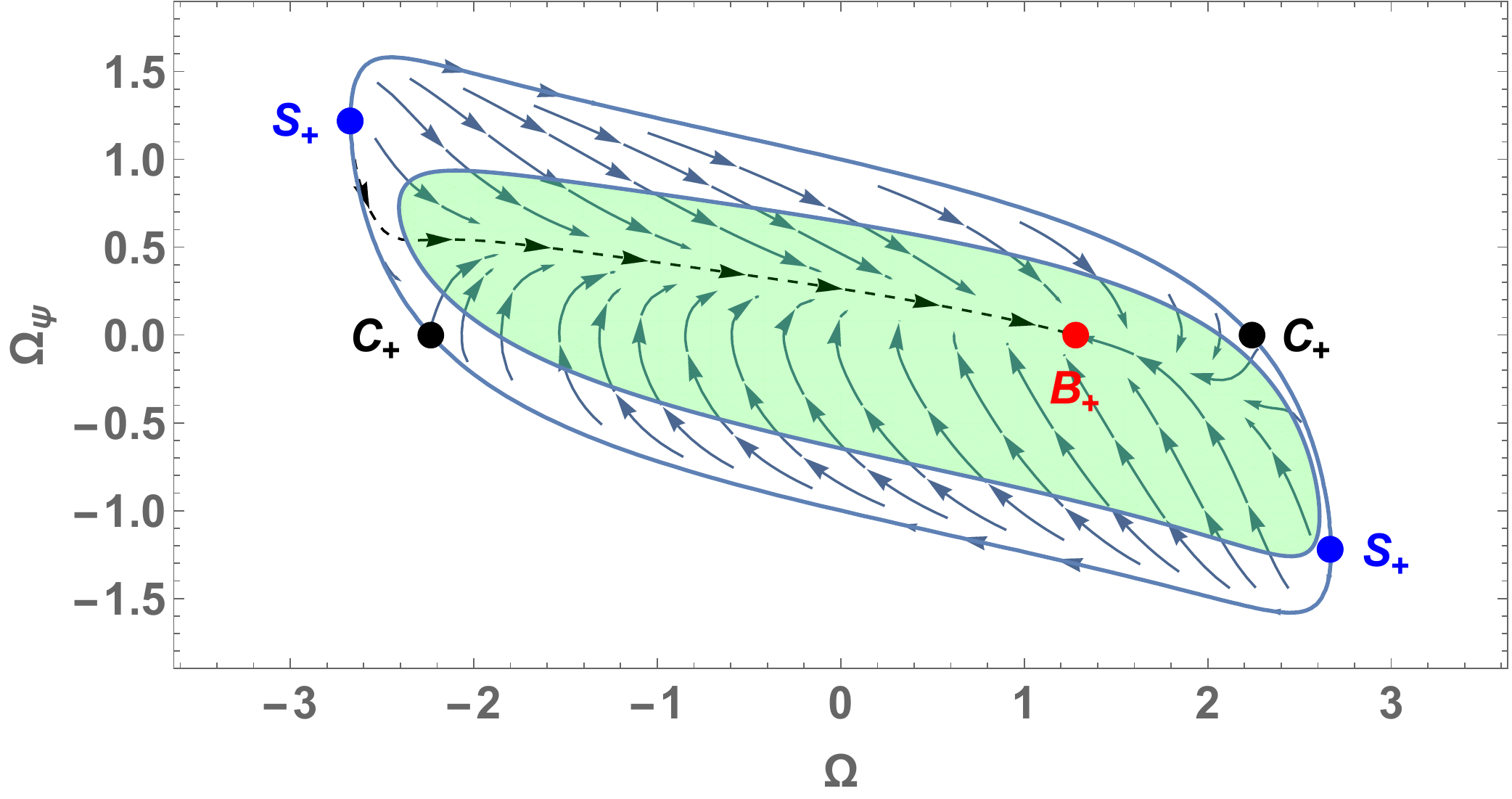}
        \includegraphics[width=0.4\textwidth]{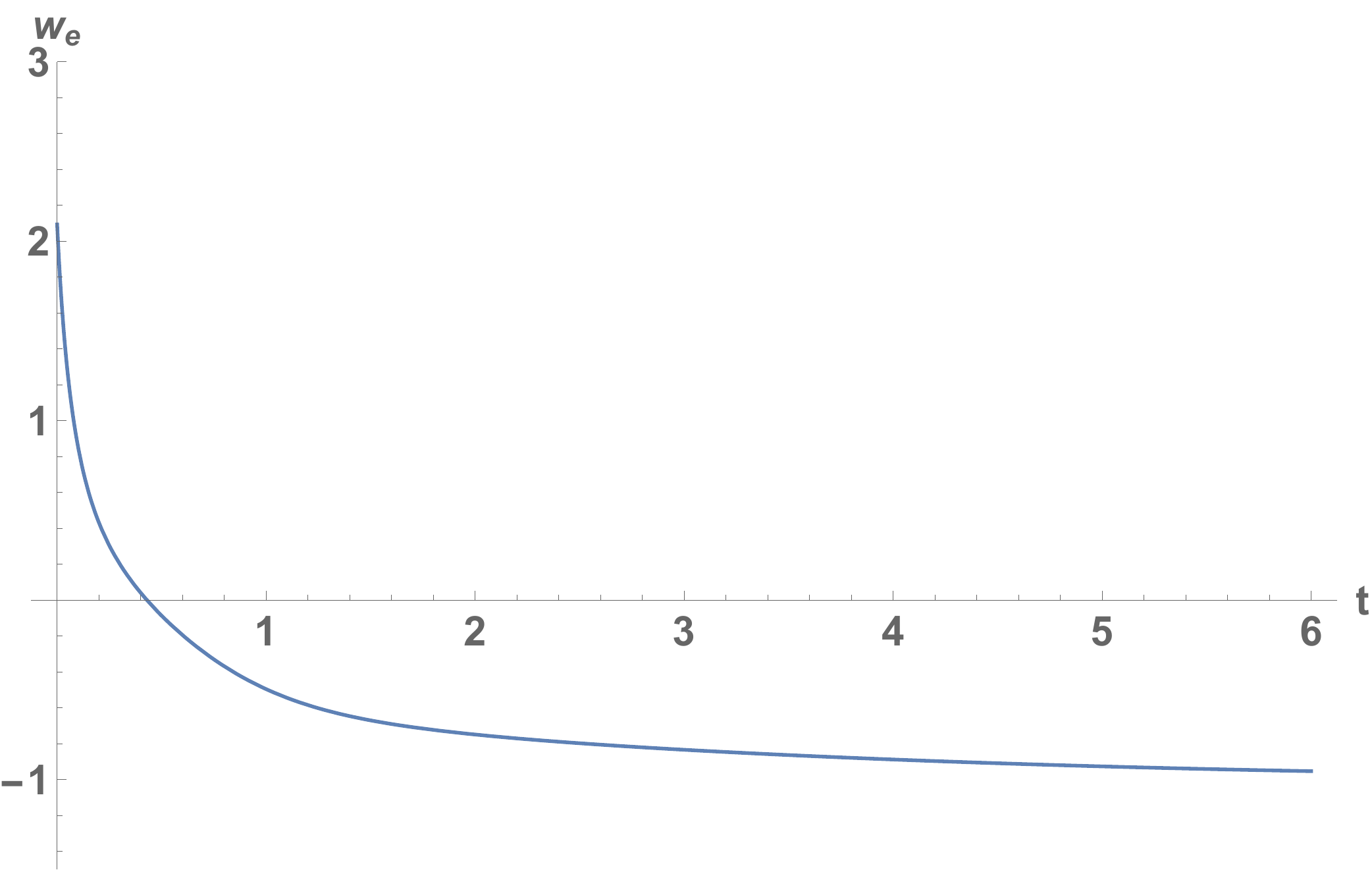}}
    \caption{{\it Top panel:} invariant subset $\Omega_H=1$ for  $\Gamma=3/4$ ($V \propto \psi^4$), $c_1=0$ and $\xi=1/10$. {\it Middle panel:} invariant subset $\Omega_H=1$ for $\Gamma=3/4$, $c_1=\sqrt{\frac{(1-\Gamma)^2}{\xi}}+1$ and $\xi=1/10$. {\it Bottom panel:} evolution of the effective equation of state parameter for the black-dashed trajectory showed in the middle panel, with initial conditions $\{ \Omega=-2.63 , \Omega_\psi = 0.9 \}$.}
     \label{fig:gamma3o4}
\end{figure}

For the massless case $\Gamma=3/4$ we give a couple of examples in Fig.~\ref{fig:gamma3o4}. In the top panel we show a case where the $\Omega_s$ splits the invariant subset in two parts. The flow of the stream plot indicates that the trajectories oscillate around the $\Omega_s$ segment.  However, this interpretation is ambiguous since the flow has to reach the invariant subset $\Omega_V=0$ to pass from one side to the other. The problem stems from our choice of variables which makes the system singular around the minimum of the potential. Thus, the cases we can interpret clearly in the range $1/2<\Gamma<1$ are the ones for which $\Omega_s$ lies outside the Friedman constraint: such a case is shown in the middle panel of Fig.~\ref{fig:gamma3o4}. For one of the trajectories of the middle panel (dashed black line) we provide also the evolution of the effective equation of state, which starts from ultra-stiff close to $S_+$ and ends up at the de Sitter sink.  

\section{Conclusions}\label{sec:Concl}

 We have started our study in a very general setup of non-minimally coupled real scalar fields in a FRW spacetime in the absence of regular matter. Namely, in a spatially curved FRW we have specified only the coupling term and not the potential of the scalar field, which we have just demanded to be positive. Transforming properly the variables of the system we have achieved to end up with a new set of dimensionless variables, which are bounded for most of the parameter ranges we consider and are well--defined even for recollapsing scenarios. In this general setup, we have investigated the general features of the system, some of which we recall below:
 \begin{itemize}
     \item There are singularities lying on the the boundaries of the Friedmann constraint. In the case of the flat spacetime $\Omega_H=\pm 1$ we have named them $S_\pm$ respectively, $S_+$ singularities act as sources and $S_-$ as sinks.
     \item For the positive curvature and the flat FRW cases the invariant subsets are compact in our new variables in the range $\xi\in(0,1/6)$.
     \item The critical points, we have found, can be separated into three categories: de Sitter points, radiation--like points and Milne--like points. The critical points of the first two categories exist for the spatially flat FRW, while the Milne--like points exist for the FRW with negative curvature. The critical points found for the spatially flat case are in agreement with those found in \cite{Hrycyna2010}. Note, however, that our analysis covers a broader family of potentials than the one in \cite{Hrycyna2010} and takes into account also collapsing scenarios. The critical points found for the negative curvature were analysed in this context for the first time.   
 \end{itemize}
 
We start the second part with the reasonable assumption assumption that $\Omega_{\partial V}$ depends only on $\Omega$ (because $V=V(\psi)$) and derive the general Eq.~\eqref{eq:gendiff}, which can be integrated in order to obtain classes of potentials.  We further specialize our investigation for spatially flat cases and constant tracking parameter $\Gamma$: on one hand, the case $\Gamma=1$ corresponds to the well-known exponential potential; on the other hand, $\Gamma\neq1$ provides the broad class of potentials given in Eq.~\eqref{eq:fun_V}.  The latter case is further divided into two main subclasses: $\Gamma>1$ corresponds to runaway potentials, while $\Gamma<1$ corresponds to potentials with positive powers of the field.  The free parameters of the model are $\xi$, $\Gamma$ and the integration constant $c_1$: we analyse in detail the ranges of values for which the de Sitter critical points exist inside the Friedmann constraint.  We find that, while the $\Gamma>1$ cases are easily interpreted as transitions from an early-time ultra-stiff era towards a late-time de Sitter expansion (possibly passing through an intermediate radiation epoch), the cases $1/2<\Gamma<1$ might present a singular behaviour introduced by our choice of coordinates; if $\Gamma<1/2$, the potentials might be real and positive only in some ranges of the field, which implies that only some portions of the parameter space can have physical interpretation.

\begin{acknowledgments}
G.A. is supported by Grant No. GA\v{C}R-17-16260Y of the Czech Science Foundation. G.L-G is supported by Grant No. GA\v{C}R-17-06962Y of the Czech Science Foundation. We would like to thank Sante Carloni and Ji\v{r}\'{i} Bi\v{c}\'{a}k  for his advises and remarks.

\end{acknowledgments}  

\bibliographystyle{unsrt}
\bibliography{refs}

\end{document}